\documentclass[12pt]{article}
\usepackage{latexsym,epsfig,amssymb,amsmath}
\usepackage{color}
\newcommand{\ft}[2]{{\textstyle\frac{#1}{#2}}}

\def\bfone{\relax{\rm 1\kern-.35em 1}}

\newcommand{\be}{\begin{equation}}
\newcommand{\ee}{\end{equation}}
\newcommand{\ben}{\begin{displaymath}}
\newcommand{\een}{\end{displaymath}}
\newcommand{\bea}{\begin{eqnarray}}
\newcommand{\eea}{\end{eqnarray}}

\newcommand{\non}{\nonumber\\}
\newcommand{\bean}{\begin{eqnarray*}}
\newcommand{\eean}{\end{eqnarray*}}
\newcommand{\beqs}{\begin{eqnarray}}
\newcommand{\eeqs}{\end{eqnarray}}

\newcommand{\mathon}{\mathversion{bold}}
\newcommand{\mathoff}{\mathversion{normal}}

\makeatletter
\@addtoreset{equation}{section}
\makeatother

\begin{document}

\thispagestyle{empty}

\begin{flushright}\small
\end{flushright}

\bigskip
\bigskip

\mathon
\vskip 10mm
\begin{center}
  {\LARGE {\bf SO(9) supergravity in two dimensions}}
\end{center}
\mathoff


\vskip 8mm

\begin{center}
{\bf Thomas Ortiz, Henning Samtleben\footnote{Institut Universitaire de France}}\\[.4ex]
{\small Universit\'e de Lyon, Laboratoire de Physique, UMR 5672, CNRS et ENS de Lyon,\\
46 all\'ee d'Italie, F-69364 Lyon CEDEX 07, France \\
{\tt thomas.ortiz , henning.samtleben , @ens-lyon.fr}}
\end{center}

\vskip2cm
\begin{center} {\bf Abstract } \end{center}
\begin{quotation}\noindent
We present maximal supergravity in two dimensions with gauge group~$SO(9)$.
The construction is based on selecting the proper embedding of the gauge group into the
infinite-dimensional symmetry group of the ungauged theory.
The bosonic part of the Lagrangian is given by a (dilaton-)gravity coupled 
non-linear gauged $\sigma$-model with Wess-Zumino term. 
We give explicit expressions for the fermionic sector, the Yukawa couplings and the scalar potential
which supports a half-supersymmetric domain wall solution.
The theory is expected to describe the low-energy effective action upon reduction
on the D0-brane near-horizon warped $AdS_2 \times S^8$ geometry,
dual to the supersymmetric (BFSS) matrix quantum mechanics.

\end{quotation}

\vfill

October 2012

\newpage
\setcounter{page}{1}

\tableofcontents

\bigskip
\bigskip


\section{Introduction}


Since the discovery of the celebrated AdS/CFT correspondence~\cite{Maldacena:1997re},
inspired by the properties of D3-branes,
various scenarios of more general gauge/gravity dualities have been put forward.
In particular, the original proposal has soon been extended to the case of 
non-conformal D$p$-branes~\cite{Itzhaki:1998dd}. 
The dual boundary theory in this case is the maximally supersymmetric $(p+1)$-dimensional 
Yang-Mills theory (non-conformal for $p\not=3$).
Relatively few tests of these non-conformal dualities have been performed
(see however~\cite{Hashimoto:1999xu,Sekino:1999av,Sekino:2000mg,Hiller:2000nf,Gherghetta:2001iv,Morales:2002ys,Asano:2004vj,Hiller:2005vf}),
and only more recently the techniques of holographic renormalization  
have been developed systematically also for 
the non-conformal case~\cite{Wiseman:2008qa,Kanitscheider:2008kd}.
Of particular interest are the dualities for low values of $p$ for which
lattice results on the field theory side 
allow to formulate and perform concrete tests of the 
correspondence~\cite{Anagnostopoulos:2007fw,Catterall:2008yz,Catterall:2010fx,Hanada:2011fq}.
In particular, for $p=0$, which is the case of interest in this paper, 
the dual field theory is the supersymmetric matrix quantum mechanics
which itself has been proposed as a non-perturbative definition of
M-theory~\cite{Banks:1996vh}.

In the supergravity approximation, the non-conformal dualities imply a 
correspondence between gauged supergravities supporting 
domain-wall solutions, and the non-conformal quantum field theories living on the domain 
walls~\cite{Boonstra:1998mp,Behrndt:1999mk}. For the $p=0$ case, the expected supergravity
theory is a two-dimensional theory with maximal supersymmetry and gauge group $SO(9)$
describing the low-dimensional excitations around the D0-brane near-horizon geometry of a 
warped product AdS$_2 \times S^8$.
Unlike all the other maximal supergravities relevant for the higher-dimensional holographic
dualities~(see e.g.~\cite{Boonstra:1998mp,Bergshoeff:2004nq} for discussion and references),
the construction of this two-dimensional theory has remained unaccomplished so far.

The aim of the present paper is precisely to fill this gap and to construct the maximally supersymmetric
two-dimensional supergravity with $SO(9)$ gauge group. The structure of two-dimensional
maximal supergravity is particularly rich. In its ungauged form, the field equations describe
a dilaton-gravity coupled non-linear $\sigma$-model with target space $E_{8(8)}/SO(16)$.
These equations are classically integrable, which leads to the existence of a linear system 
\cite{Nicolai:1987kz,Nicolai:1988jb} and an infinite-dimensional global symmetry group
$E_{9(9)}$~\cite{Julia:1982gx}, the centrally extended affine $E_{8(8)}$, which extends the target space isometries
and can be realized on-shell on the equations of motion.
As in higher dimensions, particular subgroups of the global symmetry group can be gauged preserving maximal supersymmetry
by introducing proper gauge couplings. The bosonic part of this construction has been given
in~\cite{Samtleben:2007an} with the possible gaugings parametrized by a constant embedding tensor
transforming in the (infinite-dimensional) basic representation of~$E_{9(9)}$. The fermionic sector (and the scalar potential)
have not yet been worked out in full generality, but we certainly expect that every bosonic deformation
of~\cite{Samtleben:2007an} can be consistently supersymmetrized.

In this paper, we present the detailed construction of the supergravity relevant for the D0-brane near-horizon geometry 
including its full fermionic sector and scalar potential. This theory corresponds to the gauging 
of an $SO(9)$ subgroup which however includes generators living beyond the $E_{8(8)}$ zero modes of the affine algebra.
In other words, this theory includes the gauging of symmetries beyond the standard $E_{8(8)}$ target space isometries.
Alternatively, and this is the construction we give in this paper, one may start from a T-dual version of the ungauged
theory, in which the $E_{8(8)}/SO(16)$ target space is replaced by another target space 
$\left(SL(9)\ltimes {\mathbb{T}}_{84}\right)/SO(9)$ together with a Wess-Zumino term, see the main text for details.
In this dual frame, the $SO(9)$ gauged theory can be obtained straightforwardly by gauging purely off-shell symmetries
via minimal couplings of two-dimensional (non-propagating) vector fields.
We present the explicit construction and show that the resulting theory is maximally supersymmetric when these minimal
couplings are accompanied by the proper Yukawa couplings and a scalar potential.

The paper is organized as follows. In section~\ref{sec:maximal}, we review the structure of
ungauged maximal supergravity in two dimensions. In particular, we give two different 
on-shell equivalent versions of the theory, obtained by torus reduction from three and eleven 
dimensions, respectively. In section~\ref{sec:vector}, we review the general structure of vector
fields and gaugings in two dimensions. Vector fields transform in the basic representation of the
affine $E_{9(9)}$ symmetry of the ungauged theory. They can be coupled in order 
to gauge part of the global symmetries, provided the couplings are parametrized by a constant embedding tensor which itself transforms in the basic representation of $E_{9(9)}$. 
Within this infinite-dimensional representation we identify the 36 vectors 
relevant for gauging $SO(9)$ and determine their supersymmetry transformations.
In section~\ref{sec:so9}, we perform the explicit construction of the
$SO(9)$ gauged theory by introducing minimal couplings to vector fields, Yukawa couplings in
the fermionic sector, and a scalar potential. We give the complete Lagrangian and check that
the supersymmetry algebra consistently closes on the bosonic fields.
Section~\ref{sec:properties} discusses some properties of the resulting theory. 
In particular, we show that the model admits a domain wall solution that breaks half
of the supersymmetries in accordance with its higher-dimensional interpretation.
Finally, the appendix collects our conventions and some technical parts of the calculation
of the Yukawa tensors, relegated from section~\ref{sec:so9}.

%
\section{Maximal supergravity in two dimensions}
\label{sec:maximal}
%

In this section, we review two-dimensional ungauged maximal (i.e.\ $N=16$) supergravity
which has a particularly rich structure.
In two dimensions, all bosonic degrees of freedom reside within the scalar sector
whose dynamics is described by a dilaton-coupled non-linear 
$\sigma$-model with target space $E_{8(8)}/SO(16)$.
The known integrable structure underlying the reduction of four-dimensional
Einstein gravity to two dimensions~\cite{Geroch:1970nt,Belinsky:1971nt,Maison:1978es,Julia:1981wc,Korotkin:1997fi}
extends to maximal supergravity \cite{Julia:1982gx,Nicolai:1987kz,Nicolai:1988jb,Nicolai:1998gi}.
Its equations of motion admit an infinite number of conserved charges that generate 
an infinite-dimensional global symmetry group.
For maximal supergravity, this group is $E_{9(9)}$, the centrally extended affine extension of the
$E_{8(8)}$ target space isometries.
In particular, this symmetry gives rise to an infinite tower of scalar fields which are 
related by on-shell duality equations. All of them can be determined in
terms of the `physical scalars' parametrizing the Lagrangian, 
by recursively integrating the first-order duality equations.
The linear action of the affine symmetry on the infinite set of scalar fields then
becomes a non-linear and non-local on-shell symmetry when projected down
to the physical scalars.

Supergravity theories in higher dimensions typically admit 
various on-shell equivalent formulations with different off-shell field content
related by on-shell dualities~\cite{Cremmer:1997ct}. In two dimensions,
the different off-shell formulations of maximal supergravity are described by
$\sigma$-models with different target-space geometry and Wess-Zumino term,
related by T-duality~\cite{Buscher:1987sk,Hull:1989jk,delaOssa:1992vc}.
Within the $E_{9(9)}$ picture, they correspond to choosing different sets of 
`physical scalars' within the infinite tower of scalar fields.
An explicit example for different frames of maximal supergravity in two dimensions
has been worked out in~\cite{Fre:2005si}.

In the following, we give two on-shell equivalent formulations of maximal supergravity
in two dimensions which we refer to as the `$E_{8(8)}$ frame' and the `$SL(9)$ frame',
respectively . The former one is reviewed in section~\ref{subsec:e8} and corresponds to 
the most compact form of the maximal theory. It is obtained by dimensional reduction 
from three dimensions and described by an $E_{8(8)}/SO(16)$ coset space $\sigma$-model.
In section~\ref{subsec:sl9} we construct the two-dimensional
theory which is obtained by dimensional reduction of the eleven-dimensional 
theory~\cite{Cremmer:1978km}. In this $SL(9)$ frame, the theory is described by a $\sigma$-model with
target space $\left(SL(9)\ltimes {\mathbb{T}}_{84}\right)/SO(9)$ and Wess-Zumino term.
The latter formulation will be particularly useful for the construction of the $SO(9)$ gauged 
theory in the rest of the paper, 
as the relevant $SO(9)$ gauge group can be identified 
within the $SL(9)$ target-space isometries.
In short, there are two inequivalent embeddings of $SO(9)$ into the affine $E_{9(9)}$
via
\bea
&& SO(9) ~\subset~ SL(9) ~\subset~ E_{8(8)} ~\subset~ E_{9(9)} \;,
\nonumber\\
\mbox{and}
&& SO(9) ~\subset~ SL(9) ~\subset~ \widehat{SL(9)} ~\subset~ E_{9(9)} \;.
\label{soso}
\eea
These two $SO(9)$ groups are embedded into the target-space isometries of the two frames, respectively.
It turns out~\cite{Samtleben:2007an} that the second $SO(9)$ is the relevant gauge group for the model
we are interested in (in contrast, the first one can not even consistently be gauged~\cite{Nicolai:2001sv}).
Consequently, the model is most conveniently constructed as a gauging of off-shell symmetries in the 
$SL(9)$ frame.
In preparation for the gauging, 
we discuss in section~\ref{subsec:general} in detail the structure of off-shell symmetries in the $SL(9)$ frame
and show how they embed into the full affine $E_{9(9)}$ group of on-shell symmetries.

\subsection{Reduction from three dimensions: the $E_{8(8)}$ frame}
\label{subsec:e8}

The most compact formulation of maximal supergravity in two dimensions is obtained by
dimensional reduction of the maximal three-dimensional theory \cite{Marcus:1983hb}. In this formulation
the maximal off-shell symmetry group $E_{8(8)}$ is inherited directly from the three-dimensional theory.
The bosonic sector of the theory is a (dilaton-)gravity coupled non-linear $\sigma$- model with
target space $E_{8(8)}/SO(16)$. I.e.\ the scalar fields parametrize 
$E_{8(8)}$ group-valued matrices ${\cal V}$ giving rise to the currents
\bea
{\cal V}^{-1}\,\partial_\mu {\cal V} &=& \frac12 Q_\mu^{IJ}\,X^{IJ} + P_\mu^A\,Y^A
\;,
\label{PQ}
\eea
with $X^{IJ}=X^{[IJ]}$ and $Y^A$ denoting the 120 compact and 128 non-compact generators of $E_{8(8)}$, respectively,
see e.g.~\cite{Nicolai:1998gi} for the corresponding algebra conventions.
Up to quartic fermions, the Lagrangian of two-dimensional maximal supergravity is given by%
\footnote{Our space-time signature is $(+-)$
with two-dimensional gamma-matrices satisfying the algebra $\gamma_\alpha \gamma_\beta = \eta_{\alpha\beta}+\varepsilon_{\alpha\beta}\gamma^3$\,.
}

\bea
e^{-1}\,{\cal L}_0 &=& -\frac{1}{4} \rho  R^{(2)} + \frac{1}{4} \rho \, P^{\mu A}P_{\mu}^{A}
- \rho  \,e^{-1} \varepsilon^{\mu \nu} \bar{\psi}_{2}^{I} D_{\mu} \psi_{\nu}^{I} 
-\frac{i }{2}\left( \partial^{\mu} \rho \right) \bar{\psi}_{\nu}^{I} \gamma^{\nu}  \psi_{\mu}^{I}
\nonumber\\
&&{} 
- \frac{i}{2} \, \rho \,  \bar{\chi}^{\dot{A}} \gamma^{\mu} D_{\mu} \chi^{\dot{A}} 
 -\frac{1}{2} \, \rho\,   \bar{\chi}^{\dot{A}} \gamma^{\nu}  \gamma^{\mu} \psi_{\nu}^{I} \,\Gamma_{A \dot{A}}^{I} P_{\mu}^{A}
 -\frac{i}{2} \,\rho \,  \bar{\chi}^{\dot{A}} \gamma^{3}  \gamma^{\mu} \psi_{2}^{I} \,\Gamma_{A \dot{A}}^{I} P_{\mu}^{A}
 \;,\;\;
 \label{L0E8}
\eea
with curvature scalar $R^{(2)}$, the dilaton field $\rho$, and fermions $\psi_\mu^I$, $\psi_2^I$, and $\chi^{\dot{A}}$, transforming 
in the ${\bf 16}$ and ${\bf 128}_c$ of the R-symmetry group $SO(16)$, respectively.
The two-dimensional Levi-Civita tensor density is denoted by $\varepsilon_{\mu\nu}$,
and $e\equiv\sqrt{|{\rm det}\,g_{\mu\nu}|}$ is the determinant of the two-dimensional vielbein $e_\mu{}^\alpha$.
Covariant derivatives on the fermionic fields are defined with the $SO(16)$ composite connection $Q_\mu^{IJ}$
from (\ref{PQ})
\bea
D_\mu \psi_\nu^I &:=& 
\partial_\mu \psi_\nu^I + 
\frac14\omega_\mu{}^{\alpha\beta}\,\gamma_{\alpha\beta}\, \psi_\nu^I  +
Q_\mu^{IJ} \psi_\nu^J 
\;,\non
D_\mu \chi^{\dot{A}} &:=& \partial_\mu \chi^{\dot{A}} + 
\frac14\omega_\mu{}^{\alpha\beta}\,\gamma_{\alpha\beta}\, \chi^{\dot{A}}  +
\frac14 Q_\mu^{IJ} \Gamma^{IJ}_{{\dot{A}}{\dot{B}}} \,\chi^{\dot{B}} \;,
\eea
etc., with spin connection $\omega_\mu{}^{\alpha\beta}$ and $SO(16)$ gamma matrices $\Gamma_{A \dot{A}}^{I}$.
The Lagrangian (\ref{L0E8}) is invariant under the supersymmetry transformations%
\begin{align}
\delta_{\epsilon} e_{\mu}{}^{\alpha} &= i \, \bar{\epsilon}^{I} \gamma^{\alpha} \psi_{\mu}^{I}\;,
&\delta_{\epsilon} \psi_{\mu}^{I} &= D_{\mu} \epsilon^{I}
\;,\nonumber\\
\delta_{\epsilon} \rho &= - \rho \, \bar{\epsilon}^{I} \gamma^{3} \psi_{2}^{I}\;,
&\delta_{\epsilon} \psi_{2}^{I} &= - \frac{i}{2} \, \gamma^{3}\gamma^{\mu} \epsilon^{I}\,\rho^{-1} \partial_{\mu} \rho
\;,\nonumber\\
\delta_{\epsilon} {\cal V} &= \bar{\epsilon}^{K} \Gamma_{A \dot{A}}^{K} \chi^{\dot{A}} \; ({\cal V} \,Y^A )\;,
&\delta_{\epsilon} \chi^{\dot{A}}&= \frac{i}{2} \, \Gamma_{A \dot{A}}^{I}  \gamma^{\mu}  \epsilon^{I} P_{\mu}^{A}
\;,
\end{align}
up to total derivatives.
The global off-shell symmetry $E_{8(8)}$ acts by left multiplication on the matrices ${\cal V}$ and gives rise
to the algebra-valued conserved Noether current
\bea
J_\mu &\equiv& \rho\,P_\mu^A\, ({\cal V} \,Y^A {\cal V}^{-1} )
\;.
\eea
As usual, in two dimensions such a current gives rise to the definition of 
($\mathfrak{e}_{8(8)}$-valued) dual scalar fields $Y$ according to
\bea
 \partial_\mu Y &=& -e\varepsilon_{\mu\nu} J^\nu 
 \;,
    \label{dualscalarE8}
\eea
which are the lowest members of an infinite hierarchy of dual potentials
that exhibit the integrable structure underlying the classical theory.

\subsection{Reduction from eleven dimensions: the $SL(9)$ frame}
\label{subsec:sl9}

Another formulation of the two-dimensional maximal supergravity which will turn out to
be relevant for the constructions of this paper is obtained by direct dimensional
reduction of the eleven-dimensional theory~\cite{Cremmer:1978km}.
This formulation exhibits manifest and off-shell $SL(9)$ symmetry which descends
from linearized diffeomorphisms on the nine-dimensional internal torus. 
As discussed in the introduction of this section, this $SL(9)$
is not a subgroup of the group $E_{8(8)}$ discussed above. 
Rather they both sit as subgroups in the affine extension
$E_{9(9)}\equiv\widehat{E}_{8(8)}$ with a common intersection of $SL(8)$.
In this second formulation, the bosonic sector is a (dilaton-)gravity coupled non-linear $\sigma$-model with
target space $\left(SL(9)\ltimes {\mathbb{T}}_{84}\right)/SO(9)$ and topological term. 
Its isometries are spanned by the semi-direct product of $SL(9)$ with 84 nilpotent translations ${\mathbb{T}}_{84}$
transforming in an irreducible representation of $SL(9)$.
The part of the scalar fields descending from the internal part of the eleven-dimensional metric parametrizes
$SL(9)$-valued matrices ${\cal V}_m{}^{a}$ giving rise to the currents
\bea
\left({\cal V}^{-1}\,\partial_\mu {\cal V}\right)^{ab} &=& Q_\mu^{[ab]} + P_\mu^{(ab)}
\;,
\label{coset}
\eea
of which again the $Q_\mu^{[ab]}$ play the role of $\,\mathfrak{so}(9)$ connections. The remaining
84 scalar fields which originate from the internal components of the eleven-dimensional three-form
are labeled as $\phi^{klm}=\phi^{[klm]}$ and mainly enter the Lagrangian via the currents\footnote{
In our conventions for this paper, we reserve letters $a, b, c, \dots$ from the beginning of the alphabet for `flat' $SO(9)$ indices
which are raised and lowered with $\delta_{ab}$. In contrast, the letters
$k, l, m, \dots$  indicate $SL(9)$ vector indices which transform under the global $SL(9)$ of the ungauged theory.}
\bea
\varphi^{abc}_\mu &\equiv& {\cal V}_{[klm]}{}^{abc}\,\partial_\mu\phi^{klm}
\;.
\label{curr_phi}
\eea
Here, and in the following we use the notation ${\cal V}_{[klm]}{}^{abc}\equiv {\cal V}_{[k}{}^{a}{\cal V}_l{}^{b}{\cal V}_{m]}{}^{c}$,
etc., for the group-valued $SL(9)$ matrix evaluated on tensor products.
Up to quartic fermions, the Lagrangian in this frame is given by
\bea
e^{-1}{\cal L}_0 &=& -\frac{1}{4} \rho R^{(2)}  
+ \frac{1}{4} \rho  \, P^{\mu \,ab }P_{\mu}^{ab}
+ \frac1{12}\, \rho^{1/3}  \, \varphi^{\mu \,abc}\varphi_{\mu}^{ abc} 
\nonumber\\
&&{}  {+} \frac1{648}\, e^{-1} \varepsilon^{\mu \nu} \varepsilon_{klmnpqrst}\, 
\phi^{klm}  \,\partial_{\mu} \phi^{npq} \, \partial_{\nu} \phi^{rst}  
\nonumber\\
&&{}
- \rho e^{-1}  \varepsilon^{\mu \nu} \bar{\psi}_{2}^{I} D_\mu \psi_{\nu}^{I} 
-\frac{i}{2}\, \bar{\psi}_{\nu}^{I} \gamma^{\nu}  \psi_{\mu}^{I}\,\partial^{\mu} \rho
- \frac{i}{2}\, \rho  \, \bar{\chi}^{a I} \gamma^{\mu} D_\mu \chi^{a I} 
\nonumber\\
&&{}
 -\frac{1}{2}  \,\rho  \, \bar{\chi}^{a I} \gamma^{\nu}  \gamma^{\mu} \psi_{\nu}^{J} \Gamma_{I J}^{b} P_{\mu}^{ab}
 -\frac{i}{2}  \,\rho \, \bar{\chi}^{a I} \gamma^{3}  \gamma^{\mu} \psi_{2}^{J} \Gamma_{I J}^{b} P_{\mu}^{ab}
 \nonumber\\
&&{}  - \frac14\,  \rho^{2/3}  \, \bar{\chi}^{a I}  \gamma^{3} \gamma^{\nu} \gamma^{\mu} \psi_{\nu}^{J} \Gamma_{I J}^{bc} \,\varphi_{\mu}^{ abc}
 -\frac{i}{12}\,  {\rho}^{2/3} \, \bar{\chi}^{a I} \gamma^{\mu} \psi_{2}^{J} \Gamma_{I J}^{bc} \,\varphi_{\mu}^{abc }
 \nonumber\\
&&{}
+\frac{i}{54}  \, \rho^{2/3}  \, \bar{\psi}_{2}^{I} \gamma^{3} \gamma^{\mu} \psi_{2}^{J} \Gamma_{IJ}^{abc} \,\varphi_{\mu}^{abc} 
+ \frac1{24}\, \rho^{2/3}  \, \bar{\psi}_{2}^{I} 
\Big(\gamma^{\mu} \gamma^{\nu} -\frac{1}{3} \gamma^{\nu} \gamma^{\mu}  \Big)\, \psi_{\nu}^{J} \, \Gamma_{IJ}^{abc} \,\varphi_{\mu}^{ abc}
\nonumber\\
&&{}
+ \frac{i}2\, \rho^{2/3} \, \bar{\chi}^{aI} \gamma^{3} \gamma^{\mu} \chi^{bJ} \Gamma_{IJ}^{c} \,\varphi_{\mu}^{abc} 
- \frac{i}{24}\, \rho^{2/3} \, \bar{\chi}^{aI} \gamma^{3} \gamma^{\mu} \chi^{aJ} \Gamma_{IJ}^{bcd} \,\varphi_{\mu}^{ bcd }
\;.
\label{L0SL9}
\eea
The topological (Wess-Zumino) term is defined by the totally antisymmetric $SL(9)$ tensor $\varepsilon_{klmnpqrst}$\,.
The $SO(9)$ gamma matrices are denoted by $\Gamma^a_{IJ}=\Gamma^a_{(IJ)}$ with $I, J=1, \dots, 16$.
Under $SO(9)$, the gravitino $\psi_\mu^I$ and dilatino $\psi_2^I$ transform in the ${\bf 16}$, while
the matter fermions $\chi^{aI}$ transform as a vector-spinor in the irreducible 
${\bf 128}$, i.e.\ $\Gamma^{a}_{IJ} \, \chi^{aJ} \equiv 0$\,.
Accordingly, covariant derivatives are defined as
\bea
D_{\mu} \psi_{\nu}^{I} &=& \partial_{\mu} \psi_{\nu}^{I} 
+ \frac{1}{4} {{\omega_{\mu}}^{\alpha \beta}} \gamma_{\alpha \beta} \, \psi_{\nu}^{I}
+ \frac{1}{4} Q_{\mu}^{ab} \, \Gamma_{IJ}^{ab} \, \psi_{\nu}^{J} 
\;,\nonumber\\
 D_{\mu} \chi^{aI} &=& \partial_{\mu} \chi^{aI} 
 + \frac{1}{4} {{\omega_{\mu}}^{\alpha \beta}} \gamma_{\alpha \beta} \, \chi^{aI}
 + Q_{\mu}^{ab} \, \chi^{bI}
+ \frac{1}{4} Q_{\mu}^{bc} \, \Gamma_{IJ}^{bc} \, \chi^{aJ} 
\;,
\eea
etc., with the $SO(9)$ connection $Q_{\mu}^{ab}$ from (\ref{coset}).
In principle, the Lagrangian (\ref{L0SL9}) can be obtained by explicitly performing the dimensional reduction of~\cite{Cremmer:1978km}.
Rather than going through the lengthy details of a reduction of the fermionic sector, here we have preferred to construct (\ref{L0SL9}) 
directly in two dimensions by imposing invariance under the following supersymmetry transformations
\begin{align}
\delta_{\epsilon} e_{\mu}{}^{\alpha} &= i \, \bar{\epsilon}^{I} \gamma^{\alpha} \psi_{\mu}^{I}\;,
&\delta_{\epsilon} \psi_{\mu}^{I} &= D_{\mu} \epsilon^{I} 
- \frac{1}{24}\, \rho^{-{1/3}}\, \Gamma_{IJ}^{abc} \left( \frac{1}{3} \gamma_{\mu} \gamma^{\nu} +  \gamma^{\nu} \gamma_{\mu} \right)  \gamma^{3}\epsilon^{J} \, \varphi_{\nu}^{abc}
\;,\nonumber\\
\delta_{\epsilon} \rho &= - \rho \, \bar{\epsilon}^{I} \gamma^{3} \psi_{2}^{I}\;,
&\delta_{\epsilon} \psi_{2}^{I} &= - \frac{i}{2} \, \gamma^{3}\gamma^{\mu} \epsilon^{I}\,\rho^{-1} \partial_{\mu} \rho
\;,\nonumber\\
\delta_{\epsilon} {{\cal{V}}_{i}}^{a}&= \bar{\epsilon}^{I} \Gamma_{I J}^{(a} \chi^{b)J} {\cal V}_i{}^b
\;,
&
\delta_{\epsilon} \chi^{a I}&= 
\frac{i}{2}\,  \Gamma_{IJ}^{b}  \, \gamma^{\mu}\epsilon^{J} P_{\mu}^{\left( ab\right)}  
-\frac{i}6\, \rho^{-{1/3}}   \Big( \delta^{ab}\Gamma_{IJ}^{cd}  -  
\frac{1}{6}\, \Gamma_{IJ}^{abcd} \Big)\, \gamma^{3} \gamma^{\mu}\epsilon^{J} \varphi_{\mu}^{bcd}
\;,
\nonumber\\[1ex]
%
\delta_{\epsilon} {\phi}^{ijk}&= 
\makebox[0in][l]{
$\displaystyle{\frac32\,  \rho^{{1}/{3}}\, {\cal{V}}^{-1}_{abc}{}^{[ijk]} \Gamma_{I J}^{ab}\,  \bar{\epsilon}^{I} \gamma^{3}   \chi^{cJ} 
+ \frac{1}{6}\,  \rho^{{1/3}} \, {\cal{V}}^{-1}_{abc}{}^{[ijk]} \Gamma_{IJ}^{abc} \,\bar{\epsilon}^{I} \psi_{2}^{J}
\;,}$}
\label{susy0}
\end{align}
which entirely determines (\ref{L0SL9}).
In turn, the transformations (\ref{susy0}) are determined by closure of the supersymmetry algebra
(we give more details on this algebra in section~\ref{subsec:susy_algebra} below).
The on-shell equivalence between the two Lagrangians (\ref{L0E8}) and (\ref{L0SL9}) can be made explicit
by identifying the 128 scalar fields that parametrize the target space of (\ref{L0SL9}) with a subset of the union 
of the scalar fields parametrizing (\ref{L0E8}) and their on-shell duals (\ref{dualscalarE8}).
Since in this paper we will exclusively be working with the second version of the theory, we do
not go into further details here.

The global off-shell symmetry $SL(9)$ of the Lagrangian~(\ref{L0SL9}) 
acts by left multiplication on the matrices ${\cal V}_m{}^a$ and matrix action on the
scalar fields $\phi^{kmn}$
\bea
\delta {\cal V}_m{}^a = \Lambda_m{}^n\,{\cal V}_n{}^a
\;,\qquad
\delta \phi^{klm} = -3\Lambda_n{}^{[k}\,\phi^{lm]n}
\;.
\label{sl9}
\eea
All other fields are left invariant.
The associated $\mathfrak{sl}_9$-valued conserved Noether current is given by
\bea
({{J_{\mu}})_{k}}^{l} &=&    \rho \,   {{\cal{V}}_{k}}^{a} P_{\mu}^{ab}  \,  {\cal{V}}^{-1}{}^{b l}  - \rho^{1/3} \, 
\Big(   {{\cal{V}}_{k}}^{a} {\cal{V}}^{-1}{}^{dl}\, \varphi^{bcd}   \varphi_{\mu}^{abc}  
- \frac{1}{9} \delta_{k}^{l}\,  \varphi^{abc}  \varphi_{\mu}^{abc} \Big)\nonumber\\
&&{}   {+} \frac1{54} \, e \varepsilon_{\mu \nu} \,   \varepsilon^{abc def ghi}
{{\cal{V}}_{k}}^{a}  {\cal{V}}^{-1}{}^{jl}\,\varphi^{bcj}  \varphi^{def} \varphi^{\nu \, ghi} ~+~{\rm fermions}
\;,
\eea
where in analogy to (\ref{curr_phi}) we have defined the dressed scalar fields 
$\varphi^{abc} \equiv {\cal V}_{[klm]}{}^{abc} \phi^{klm}$\,.
As usual in two dimensions, the existence of this conserved current 
can be employed to define dual scalar fields $Y_{k}{}^{l}$ according to
\begin{equation}
\partial_{\mu} Y_{k}{}^{l} = - e \varepsilon_{\mu \nu} \, ({J^{\nu})_{k}}^{l}
\;.
\label{dualY}
\end{equation}
These dual scalar fields which are defined on-shell, will play an important role
in the construction of the gauged theory. For later use, we note that on-shell the
supersymmetry algebra closes on the dual scalar fields provided that we impose their 
supersymmetry transformation rules to be
\bea
\delta_{\epsilon} {Y_{k}}^{l}&=&
 \bar{\chi}^{aI} \gamma^{3}\epsilon^{J} \,{{\cal{V}}_{k}}^{b}{\cal{V}}^{-1}{}^{c l} \left( 
 \frac16 \, \rho^{1/3} 
 \left(\varphi^{agh}\varphi^{efc} \delta^{db}-
 \delta^{b[a} \varphi^{gh]c}  \varphi^{def} \right)  \Gamma_{IJ}^{defgh} 
 - \rho \, \delta^{a(b} \, \Gamma_{IJ}^{c)}\right) 
\nonumber\\   
&&{} + \frac32\, \rho^{2/3} \, \bar{\chi}^{aI}\epsilon^{J}\,{\cal{V}}^{-1}{}^{gl} {{\cal{V}}_{k}}^{[a} \varphi^{bc]g} \,\Gamma_{IJ}^{bc}
+ \frac13\,\rho^{2/3} \, \bar{\psi}_{2}^{I} \gamma^{3} \epsilon^{J} \, {\cal{V}}^{-1}{}^{gl} {{\cal{V}}_{k}}^{a} \varphi^{bcg} \,\Gamma_{IJ}^{abc}   \nonumber\\
&&{} + \bar{\psi}_{2}^{I} \epsilon^{J} \left( \frac12\, \rho \, {\cal{V}}^{-1}{}^{al} {{\cal{V}}_{k}}^{b} \,\Gamma_{IJ}^{ab} 
+ \frac1{54} \, \rho^{1/3} \,  {\cal{V}}^{-1}{}^{gl} {{\cal{V}}_{k}}^{d}\,\varphi^{abc} \varphi^{efg} 
\,\Gamma_{IJ}^{abcdef}  \right)  
\;.
\label{susyY}
\eea
Namely, evaluating two supersymmetry transformations on the dual scalars, we find with (\ref{susyY})
and upon using the duality equations~(\ref{dualY}) 
\bea
{}[\delta_{\epsilon_1},\delta_{\epsilon_2}]\, {Y_{k}}^{l}
&=& - e\varepsilon_{\mu \nu} \,\xi^{\mu}\, ({J^{\nu})_{k}}^{l} ~=~
\xi^{\mu}\,\partial_{\mu} Y_{k}{}^{l}
\;,
\label{susy_algebra_Y}
\eea
i.e.\ two
supersymmetries close in the standard way into diffeomorphisms with the parameter 
$\xi^{\mu} \equiv i \, \bar{\epsilon}_{2}^{I} \gamma^{\mu} \epsilon_{1}^{I}$.
This entirely fixes the transformation (\ref{susyY}) of the dual scalar fields.

For the rest of this section, let us summarize the remaining global off-shell symmetries 
of the Lagrangian (\ref{L0SL9}).
Apart from the global $SL(9)$ of (\ref{sl9}), the 84 translations
\bea
\delta \phi^{klm}=\Lambda^{klm}
\;,
\label{off84}
\eea
leave the Lagrangian invariant up to a total derivative.
The higher-dimensional origin of these symmetries
are the eleven-dimensional tensor gauge transformations
linear in the compactified coordinates.
In analogy to (\ref{dualY}), the associated conserved Noether current $j^\mu{}_{kmn}$ 
defines dual scalars ${Y}_{kmn}$ according to
\bea
\partial_\mu {Y}_{kmn} &=& -e\varepsilon_{\mu\nu} j^\nu{}_{kmn}
\;. 
\label{dualYY}
\eea
The last remaining global off-shell symmetry is the standard two-dimensional Weyl rescaling
\bea
&&\delta_\kappa e_\mu{}^\alpha=\kappa e_\mu{}^\alpha\;,\quad
\delta_\kappa \psi^I_\mu = \frac\kappa2 \,\psi^I_\mu \;,
\nonumber\\
&&
\delta_\kappa \chi^{aI} = -\frac\kappa2 \,\chi^{aI}\;,\quad
\delta_\kappa \psi^I_2 = -\frac\kappa2 \,\psi^I_2\;,
\label{weyl}
\eea
properly extended to the fermionic fields. It leaves all scalar fields invariant.
The associated conserved Noether current
\bea
j_\mu &\equiv& \partial_\mu\rho + {\rm fermions}
\;, 
\eea
defines the dual scalar potential $\tilde\rho$ (dual dilaton) according to
\bea
\partial_\mu \tilde\rho &=& -e\varepsilon_{\mu\nu} j^\nu{}
\;. 
\label{rhotilde}
\eea

\subsection{General symmetry structure}
\label{subsec:general}

The off-shell symmetries of the Lagrangian (\ref{L0SL9}) combine the global $SL(9)$ of (\ref{sl9}) and 
the 84 translations of (\ref{off84}) and close into the semi-direct product $SL(9)\ltimes \mathbb{T}_{84}$\,.
In fact, this finite-dimensional symmetry algebra is but a tiny glimpse of the 
full symmetry of the theory: on-shell it may be extended to the full affine Kac-Moody
algebra $\mathfrak{e}_{9(9)}\equiv \widehat{\mathfrak{e}}_{8(8)}$\,.
The explicit realization of the full affine symmetry is most conveniently formulated in the $E_{8(8)}$ frame
described in section~\ref{subsec:e8}, see e.g.~\cite{Julia:1982gx,Nicolai:1988jb,Nicolai:1998gi}, 
but will not be essential for the following construction.
Rather, we restrict here to sketching the embedding of the symmetries manifest in (\ref{L0SL9}) 
into the full picture.
Under $\mathfrak{sl}_9$, the affine algebra decomposes into
\bea
\mathfrak{e}_{9(9)}&\rightarrow&
\dots\,\oplus\, {\bf 84}_{-2/3}\,\oplus\, {\bf 84}'_{-1/3}\,\oplus\,
({K}_0 \oplus {\bf 80}_{0})\,\oplus\,{\bf 84}_{+1/3}\,\oplus\, {\bf 84}'_{+2/3}\,\oplus\,{\bf 80}_{+1}\,\oplus\,
\dots\;.
\nonumber\\
\label{adjSL9}
\eea
Here, the subscripts refer to the charges under the derivation ${\bf d}$ associated with the 
affine subalgebra $\widehat{\mathfrak{sl}_9}$ according to
\bea
{}[\,T_{\alpha,m}\,,\,T_{\beta,n}\,] &=& f_{\alpha\beta}{}^{\gamma}\,T_{\gamma,m+n}+
m\,\delta_{m+n}\,\eta_{\alpha\beta}\,{ K}_0
\;,\nonumber\\
{}
[\,{\bf d} \,,T_{\alpha,m}\,]&=& -m\,T_{\alpha,m} \;,
\label{affine2}
\eea
with $\mathfrak{sl}_9$ adjoint indices $\alpha, \beta, \dots$\,, structure constants $f_{\alpha\beta}{}^{\gamma}$\,,
Cartan-Killing form $\eta_{\alpha\beta}$,
and the generators $T_{\alpha,m}$ corresponding to ${\bf 80}_m$ in the decomposition (\ref{adjSL9}).
The $SL(9)\ltimes \mathbb{T}_{84}$ off-shell symmetries of the Lagrangian (\ref{L0SL9}) correspond to
the generators ${\bf 80}_0$ and ${\bf 84}_{+1/3}$ in (\ref{adjSL9}), while the central extension ${K}_0$
is realized~\cite{Julia:1981wc} by the extended Weyl rescaling (\ref{weyl}).
By taking successive commutators, the generators ${\bf 84}_{+1/3}$ eventually generate  the entire positive half
of the Kac-Moody algebra $\mathfrak{e}_{9(9)}$ in the grading of (\ref{adjSL9}). 
However all higher-level generators act exclusively on the
infinite tower of dual scalar potentials of which we have introduced the lowest members 
(\ref{dualY}), (\ref{dualYY}) in the previous section,
such as
\bea
{\bf 84}'_{+2/3} &:&  
\delta {Y}_{kmn} = \Lambda_{kmn}
\;,\nonumber\\
{\bf 80}_{+1} &:&  \delta {Y}_{k}{}^l = \Lambda^{(1)}{}_{k}{}^l
\;,
\label{shiftsym}
\eea
etc.. In particular, the action of the translations ${\bf 84}_{+1/3}$
commutes when evaluated on the physical fields of the Lagrangian~(\ref{L0SL9}).
More interesting are the symmetry generators of negative grading: 
they correspond to an infinite chain of `hidden' on-shell symmetries which are
realized rather non-trivially on the physical fields. Closed expressions for the action of these
generators would require to construct the analogue of the linear system of 
\cite{Maison:1978es,Julia:1981wc,Nicolai:1987kz} in the $SL(9)$ frame.
This is beyond the scope of the present paper and not relevant for the following construction.
The theory we will present later in this paper will be constructed by gauging the compact $SO(9)$ 
subgroup of the global symmetry (\ref{sl9}).

Let us finally note that the derivation ${\bf d}$, which extends the affine algebra 
$\widehat{\mathfrak{sl}_9}$ to (\ref{affine2})
is realized as an on-shell scaling symmetry of the theory 
which acts exclusively on the bosonic fields
\bea
\delta\,\rho = \lambda \,\rho\;, \qquad \delta\,\phi^{klm}=\frac \lambda3\,\phi^{klm}
\;,
\label{L0a}
\eea
and scales the Lagrangian (\ref{L0SL9}) as $\delta {\cal L}_0 = \lambda {\cal L}_0$\,. 
It extends to the dual scalar fields according to their definition
\bea
\delta\,\tilde\rho =  \lambda\,\tilde\rho\;, \qquad
\delta\,{Y}_{klm}=\frac{2\lambda}3\,{Y}_{klm}
\;, \qquad \delta\,Y_k{}^l= \lambda\,Y_k{}^l
\;,
\label{L0b}
\eea
etc..



\section{Vector fields and gauging}
\label{sec:vector}

The bosonic matter sector of two-dimensional maximal supergravity (\ref{L0SL9}) is 
built from 128 scalar fields and a dilaton. In order to gauge a subgroup of the global
symmetry group we need to introduce vector fields compatible with maximal 
supersymmetry. In this section we first discuss the general (infinite-dimensional) 
representation content of vector fields by which the two-dimensional theory can be consistently extended 
and explicitly determine the supersymmetry transformations of the lowest components
from closure of the supersymmetry algebra. We then review how particular components
of these vector fields can be employed in order to gauge the compact $SO(9)$
subgroup of (\ref{sl9}).

The reader who is merely interested in the explicit SO(9)-gauged theory 
(rather than the algebraic structure underlying general gauge deformations) 
is invited to jump directly to the last paragraph 
of this section in which we give the minimal couplings (\ref{minimalso9}) relevant for its construction.

\subsection{Vector fields and supersymmetry}
\label{subsec:vector}

In two-dimensional maximal supergravity, 
the vector fields $A_\mu{}^{\cal M}$ transform in 
the basic representation of $\mathfrak{e}_{9(9)}$~\cite{Samtleben:2007an},
i.e.\ the unique level~1 representation of the affine algebra.
Under $\mathfrak{sl}_{9}$, this representation decomposes into
\bea
{\cal R}_{{\rm vectors}}&\rightarrow&
{\bf 9}_{5/9}\,\oplus\,\nonumber\\&&{}
{\bf 36}'_{2/9}\,\oplus\,\nonumber\\&&{}
{\bf 126}_{-1/9}\,\oplus\,\nonumber\\&&{}
({\bf 9}\oplus{\bf 315}')_{-4/9}\,\oplus\,\nonumber\\&&{}
({\bf 36}'\oplus{\bf 45}'\oplus{\bf 720})_{-7/9}
\,\oplus\,\dots
\;,
\label{rep_vector}
\eea
where again subscripts refer to the charge under the derivation ${\bf d}$
of $\widehat{\mathfrak{sl}_9}$
(and the representations with equal charge modulo 1 combine into irreducible highest weight
representations of $\widehat{\mathfrak{sl}_9}$~\cite{Kac:1988iu}).
The somewhat surprising fractional charges can be confirmed by tracing back
the higher-dimensional origin of the two-dimensional vector fields. To this end, we recall 
that in the reduction of eleven-dimensional supergravity on a torus $T^9$,
the lowest level vector fields ${\bf 9}+{\bf 36'}$ in (\ref{rep_vector}) descend from the 
Kaluza-Klein vector fields $A_{\mu\,k}$ in the eleven-dimensional metric and 
the vector components $ A_{\mu}{}^{kl}$ of the antisymmetric three-form, respectively.
Starting from the standard compactification ansatz for the eleven-dimensional vielbein and 
three-form\footnote{In our conventions
we split the eleven-dimensional coordinates according to $x^M \rightarrow (x^\mu, y_m)$.}
\bea
E_{M}{}^{A} &=& 
\left(
\begin{array}{cc}
\rho^s\, e_\mu{}^\alpha & \rho^{1/9}\,A_{\mu\,k}\,{\cal V}^{-1}{}^{a\,k}
\\
0& \rho^{1/9}\,{\cal V}^{-1}{}^{a\,m}
\end{array}
\right)
\;,
\nonumber\\[1ex]
A_{MNK} &=& \left(
0\, , 0\,  , A_{\mu}{}^{mn}+A_{\mu\,k} \, \phi^{kmn} \, ,\; \phi^{kmn}
\right)
\;,
\eea
respectively, leads to a two-dimensional Lagrangian, whose pertinent terms include 
\bea
   e^{-1} {\cal L}_{2d} &=& 
   -\frac14  \rho R^{(2)}  -\left(\frac{s}2+\frac{2} {9}\right)  \rho^{-1} \partial_\mu \rho\, \partial^\mu \rho 
           - \frac1{16} \,  \rho^{11/9-2s} \, {\cal M}^{-1}{}^{kl}\, F_{\mu \nu\,k} F^{\mu \nu}{}_l 
\nonumber \\
&&{}               
-\frac1{8} \rho^{5/9-2s} 
\left(F_{\mu\nu}{}^{kl}+\phi^{klp}F_{\mu\nu\,p}\right)
\,{\cal M}{}_{km}{}{\cal M}{}_{ln}
\left(F^{\mu\nu}{}^{mn}+\phi^{mnq}F^{\mu\nu}{}_{q}\right)
\nonumber \\[1.5ex]
&&{}               
+ \dots
\;,
\label{Lag_red}
\eea
with ${\cal M}_{kl}={\cal V}_k{}^a {\cal V}_l{}^a$\,.
Elimination of the $\partial_\mu \rho\, \partial^\mu \rho$  term selects $s=-4/9$ 
and brings the Lagrangian into the frame of (\ref{L0SL9}).
Together with the fact that the entire Lagrangian carries charge $+1$ under the scaling~(\ref{L0b}),
the corresponding charges of the vector fields $A_{\mu\,k}$ and $A_{\mu}{}^{kl}$ can then be read off
from (\ref{Lag_red}).
This confirms the charge assignment of the lowest levels in~(\ref{rep_vector})
which then carries on in steps of $1/3$.

We take the occasion to derive the supersymmetry transformation of the lowest components
of vector fields which are determined by closure of the supersymmetry algebra 
(up to a global factor that can be absorbed by rescaling of the vector fields) and read\footnote{
Strictly speaking, demanding closure of the supersymmetry algebra on the lowest components
$A_{\mu\,k}$ determines their supersymmetry transformation only up to a free parameter $c$ corresponding to its
charge under the scaling~(\ref{L0b}): $\delta_{\epsilon}\,A_{\mu\,k} \propto  \rho^{c} \left( 
\bar{\psi}_{\mu}^{I} \gamma^{3} \epsilon^{J} \Gamma_{IJ}^{a}  
-  i c  \bar{\psi}_{2}^{I} \gamma_{\mu} \epsilon^{J} \Gamma_{IJ}^{a} 
-  i \bar{\chi}^{aI}  \gamma^{3} \gamma_{\mu} \epsilon^{I} \right) {{\cal{V}}_{k}}^{a} $\;.
However, the algebra uniquely fixes the supersymmetry transformation of the next components $A_{\mu}{}^{kl}$
with charge $-2/9$, thus $c=-5/9$ by comparison of the charges.
}
\bea
\delta_{\epsilon}\,A_{\mu\,k} &=&  -2 \rho^{-5/9}\Big( 
\bar{\psi}_{\mu}^{I} \gamma^{3} \epsilon^{J} \Gamma_{IJ}^{a}  
+ \frac{5i}{9}   \bar{\psi}_{2}^{I} \gamma_{\mu} \epsilon^{J} \Gamma_{IJ}^{a} 
- i \bar{\chi}^{aI}  \gamma^{3} \gamma_{\mu} \epsilon^{I} \Big) {{\cal{V}}_{k}}^{a} 
\;,
\nonumber\\[2ex]
\delta_{\epsilon}\,A_{\mu}{}^{kl} &=&   \rho^{-2/9}\Big( 
\bar{\psi}_{\mu}^{I} \epsilon^{J} \Gamma_{IJ}^{ab} 
- \frac{2i}{9}\, \bar{\psi}_{2}^{I} \gamma^{3} \gamma_{\mu} \epsilon^{J} \Gamma_{IJ}^{ab} 
 - 2 i  \, \bar{\chi}^{I[a} \gamma_{\mu} \epsilon^{J} \Gamma_{IJ}^{b]}
 \Big)
 {\cal{V}}^{-1}{}_{[ab]}{}^{kl} 
 \nonumber\\[.5ex]
&&{}
 -   {\cal{V}}^{-1}{}_{[abc]}{}^{klm}\, \varphi^{abc} \left(\delta_{\epsilon}\,A_{\mu\,m}\right)
 \;.
 \nonumber\\
 \label{susyvec}
 \eea
Since vector fields are not propagating in two dimensions
the supersymmetry algebra closes on-shell on these fields only under the condition
that their fields strengths vanish $F_{\mu\nu\,k}=0=F_{\mu\nu}{}^{kl}$.
This is the standard appearance of $(D-1)$-forms in $D$-dimensional ungauged supergravity,
see e.g.~\cite{Bergshoeff:1996ui}.
Combining~(\ref{susyvec}) with (\ref{susy0}), the supersymmetry transformations close
into gauge transformations with real parameters
\bea
\Lambda_k &=& -2\rho^{-5/9} \, \bar{\epsilon}_{1}^{I} \gamma^{3} \epsilon_{2}^{J} \Gamma_{IJ}^{a} {{\cal{V}}_{k}}^{a}
\;,
\nonumber\\[.5ex]
\Lambda^{kl} &=&  \rho^{-2/9} \, \bar{\epsilon}_{1}^{I} \epsilon_2^{J} \Gamma_{IJ}^{ab} \, {\cal{V}}^{-1}{}_{[ab]}{}^{kl}\, 
+ 2 \, \rho^{-5/9}\, \bar{\epsilon}_{1}^{I} \gamma^{3} \epsilon_{2}^{J} \Gamma_{IJ}^{a} \, 
{\cal{V}}^{-1}{}_{[bc]}{}^{kl}\,  \varphi^{abc}
\;.
\label{lamlam}
\eea
Nicely, supersymmetry thus provides another (and entirely two-dimensional)
justification for the assignment of charges 
under the scaling~(\ref{L0b}) in the vector field
representation~(\ref{rep_vector}).

\subsection{The embedding tensor in two dimensions}
\label{subsec:embedding}

Given the vector field content and the global symmetry algebra of the theory,
gaugings are most conveniently described by the embedding tensor 
formalism~\cite{Nicolai:2000sc,deWit:2002vt,deWit:2005hv}. For the two-dimensional 
supergravities with affine global symmetry algebra and vector fields in the
basic representation, the general formalism has been set up in~\cite{Samtleben:2007an}.
An arbitrary gauging is described by an {\em embedding tensor} $\Theta_{{\cal M}}{}^{{\cal A}}$
that defines the gauge group generators
\bea
X_{{\cal M}}&\equiv& \Theta_{{\cal M}}{}^{{\cal A}}\,T_{{\cal A}}
\;,
\label{genX}
\eea
as linear combinations of the generators $T_{{\cal A}}$
of the global symmetry algebra.
The algebra spanned by the generators (\ref{genX})
then is promoted to a local symmetry by introducing covariant derivatives
\bea
{\cal D}_\mu &=
\partial_\mu - g \, {\cal A}_\mu{}^{{\cal M}} \, 
\Theta_{{\cal M}}{}^{{\cal A}}\,T_{{\cal A}} \;,
\label{covD}
\eea
with gauge coupling constant $g$.
In general dimension $D$, the embedding tensor transforms in the representation
dual to the representation of $(D-1)$-forms of the theory. In two dimensions, 
with vector fields $A_\mu{}^{\cal M}$ in the basic representation (\ref{rep_vector}),
it turns out that $\Theta_{{\cal M}}{}^{{\cal A}}$ is parametrized by a constant tensor
$\Theta_{{\cal M}}$ according to
\bea
\Theta_{{\cal M}}{}^{{\cal A}}&=& 
(T_{\cal B})_{\cal M}{}^{{\cal N}}\,\eta^{{\cal A}{\cal B}}\,\Theta_{{\cal N}}
\;,
\label{ConstraintLinear}
\eea
where $\eta^{{\cal A}{\cal B}}$ denotes a particular invariant form on the affine algebra,
see~\cite{Samtleben:2007an} for details.
The lowest components of $\Theta_{{\cal M}}$ are given as
\bea
{\cal R}_{{\Theta}}&\rightarrow&
{\bf 9}'_{-14/9}\,\oplus\,\nonumber\\&&{}
{\bf 36}_{-11/9}\,\oplus\,\nonumber\\&&{}
{\bf 126}'_{-8/9}\,\oplus\,\nonumber\\&&{}
({\bf 9}'\oplus{\bf 315})_{-5/9}\,\oplus\,\nonumber\\&&{}
({\bf 36}\oplus{\bf 45}\oplus{\bf 720}')_{-2/9}
\,\oplus\,\dots
\;.
\label{rep_theta}
\eea
With respect to the full algebra (\ref{affine2}), the embedding tensor 
does not exactly transform in the representation dual to (\ref{rep_vector}) 
but comes with shifted charges with respect to the derivation ${\bf d}$.
This is a consequence of the fact that the Lagrangian itself is charged under the action of ${\bf d}$.
In order to make this explicit, we note that the equations of motion for the vector fields
arising in the reduction of~(\ref{Lag_red}) can be integrated up to
\bea
{\cal M}^{-1 kl}\,F_{\mu \nu\,l} + 2\,\rho^{-2/3} \phi^{kmn} {\cal M}_{mp} {\cal M}_{nq} \left(F_{\mu\nu}{}^{pq}+\phi^{pqr}{ F}_{\mu\nu\,r}\right)
 &=&  e\varepsilon_{\mu\nu}\,
\rho^{-19/9} \,  \theta^k\;,\nonumber\\
{\cal M}_{km} {\cal M}_{ln}\left(
F_{\mu\nu}{}^{mn}+\phi^{mnp}{F}_{\mu\nu\,p}\right)
&=&  e\varepsilon_{\mu\nu}\,
\rho^{-13/9} \, \tilde\theta_{kl}
\;,
\label{FTheta}
\eea
with integration constants $\theta^l$, $\tilde\theta_{mn}=\tilde\theta_{[mn]}$.
The ungauged theory (\ref{L0SL9}) is obtained from (\ref{Lag_red}) upon eliminating the field strengths 
from the Lagrangian by virtue of (\ref{FTheta}) with zero integration constants.
Non-trivial $\theta^l$, $\tilde\theta_{mn}$ in (\ref{FTheta}) induce a massive deformation of
the two-dimensional theory. In particular, eliminating the field strengths in this case 
gives rise to a scalar potential of the form
\bea
V_{\theta}&=&
\frac1{8} \rho^{-19/9} \left(\theta^k - 2\phi^{kmn} \tilde\theta_{mn}\right) {\cal M}_{kl} \left(\theta^k - 2\phi^{kmn} \tilde\theta_{mn}\right) 
\nonumber\\
&&{}
+\frac1{4}\,\rho^{-13/9}\,\tilde\theta_{kl}\,{\cal M}^{-1km} {\cal M}^{-1ln}\,\tilde\theta_{mn}
\;,
\label{pot_red}
\eea
quadratic in the integration constants $\theta^l$, $\tilde\theta_{mn}$. In
the embedding tensor formalism, massive deformations are treated on
the same footing as gaugings and induced by particular components of
the embedding tensor. Specifically, the integration constants $\theta^l$ and $\tilde\theta_{mn}$
are to be identified
with the lowest components of (\ref{rep_theta}). Their charges under ${\bf d}$
can thus be read off from (\ref{FTheta}) and confirm the assignment of (\ref{rep_theta}).

\begin{figure}[bt]
   \centering
   \includegraphics[width=11.6cm]{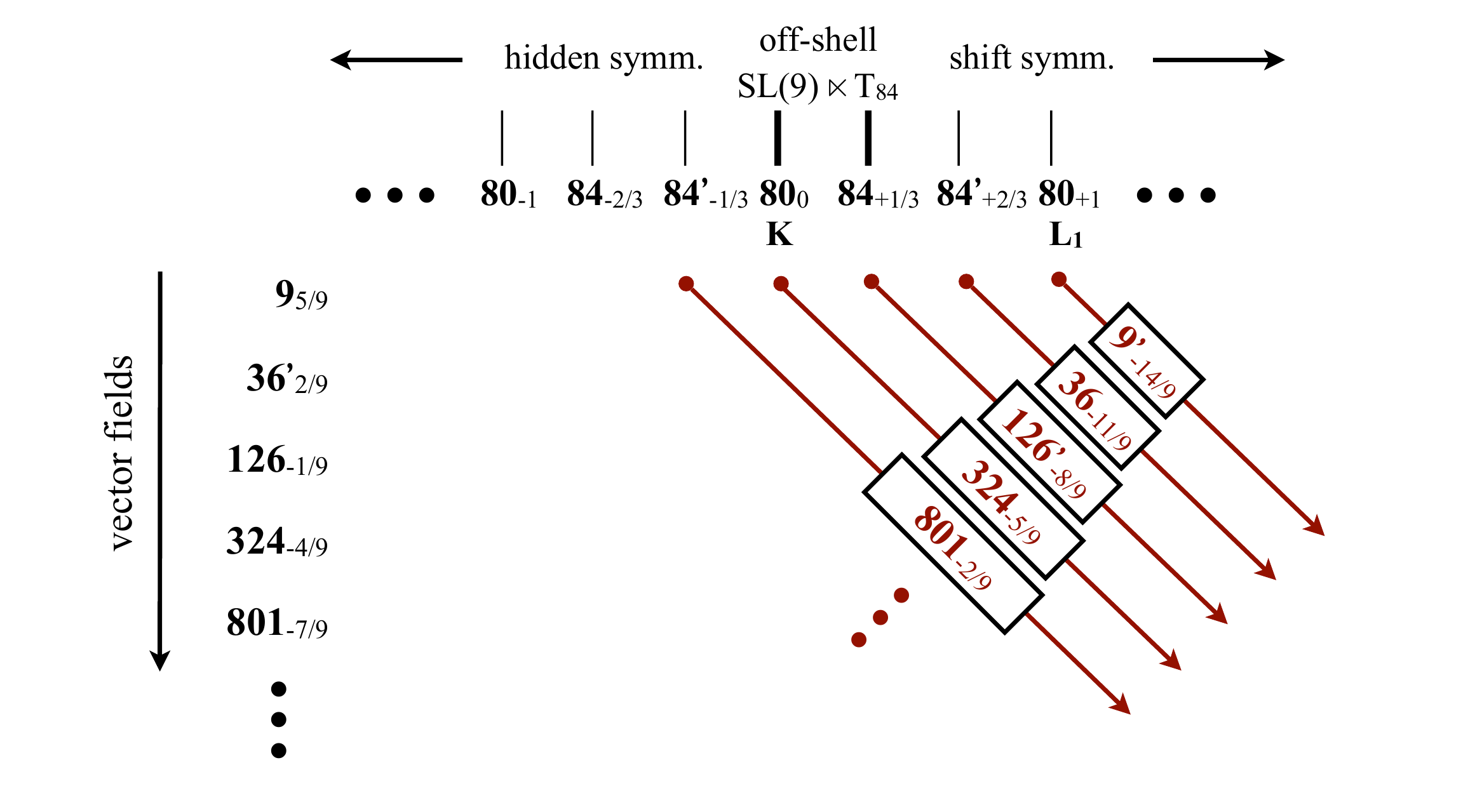}
   \caption{{\small Minimal couplings (\ref{covD}) induced by different components of the embedding tensor~$\Theta_{{\cal M}}$.}}
   \label{Figure:theta}
\end{figure}

With the representation content (\ref{rep_vector}), (\ref{rep_theta}) and the minimal couplings 
(\ref{covD}), (\ref{ConstraintLinear}), one can read off the structure of the gauge algebra,
i.e.\ identify the global symmetry generators that are promoted to local gauge symmetry.
The couplings induced by choosing an embedding tensor in a given sub-representation of
(\ref{rep_theta}) are schematically depicted in figure~\ref{Figure:theta}, organized w.r.t.\ the
respective charges under the derivation ${\bf d}$. Every diagonal line in the figure corresponds to the
components of the embedding tensor at a given charge, and indicates the induced 
couplings between vector fields and symmetry generators.
E.g.\ a gauging induced by the lowest component $\theta^k$ transforming in the ${\bf 9}'_{-14/9}$
does not trigger any gauging of the off-shell symmetries of the Lagrangian. Rather it
corresponds to the massive deformation discussed in (\ref{pot_red}) above.
On-shell however, figure~\ref{Figure:theta} shows that this deformation induces a gauging
of some of the generators in the ${\bf 80}_{+1}$ corresponding to the on-shell shift symmetries~(\ref{shiftsym})
(as well as a gauging of the Virasoro generator ${\bf L}_1$ acting as a shift on the dual dilaton (\ref{rhotilde})).

The theory which we will construct in this paper is induced by an embedding tensor $\theta_{kl}=\theta_{(kl)}$
transforming in the ${\bf 45}_{-2/9}$ from the fifth level of (\ref{rep_theta}). Figure~\ref{Figure:theta}
shows that these components induce a gauging of generators within the off-shell $SL(9)$ coupled to the vector fields $A_\mu{}^{kl}$
from (\ref{susyvec}). In contrast, none of the --- potentially possible --- generators from the ${\bf 84'}_{-1/3}$ and
${\bf 84}_{+1/3}$ are involved in this gauging, as there are forbidden by the structure of representations:
no ${\bf 45'}$ appears in the tensor products ${\bf 9}\otimes {\bf 84'}$ and ${\bf 126}\otimes {\bf 84}$ as would
be required for such couplings to take place. For the same reason, the central charge generator ${K}$
is not gauged in this theory.
More explicitly, the induced gauged subalgebra of $\mathfrak{sl}_9$ is generated by
\bea
X_{kl} &\equiv& \theta_{m[k}\,T_{l]}{}^m
\;,
\label{so9}
\eea
with the traceless $T_{k}{}^l$ denoting the generators of $\mathfrak{sl}_9$. A quick calculation confirms~\cite{deWit:2002vt}
that the resulting algebra is given by $\mathfrak{cso}_{p,q,r}$ (non-semisimple for $r>0$) with the integers $p+q+r=9$ 
characterizing the signature of $\theta_{kl}$.
In particular --- and this will be the theory of most interest in the following --- choosing $\theta_{kl}\equiv \delta_{kl}$
amounts to gauging the full compact $\mathfrak{so}(9)\subset \mathfrak{sl}_9$.
Figure~\ref{Figure:theta} indicates that the full gauge algebra of the theory is
given by an infinite-dimensional Borel subalgebra of $\mathfrak{e}_{9(9)}$ of which however
all but a finite number of generators (that correspond to the compact $\mathfrak{so}(9)$) do not act on
the physical fields and are only visible as shift symmetries on the infinite tower of dual scalar potentials,
\`a la~(\ref{shiftsym}).

In general, the gauging defined by an embedding tensor according to (\ref{covD}) is consistent only in 
case the embedding tensor satisfies an additional quadratic constraint which in two dimensions takes the form
\bea
\Theta_{{\cal M}}{}^{{\cal A}}\,\Theta_{{\cal N}}{}^{{\cal B}}\,\eta_{{\cal AB}}
~=~ 0 \qquad
\Longleftrightarrow
\qquad
\eta^{{\cal A}{\cal B}}\,T_{{\cal A},{\cal M}}{}^{{\cal P}}\,T_{{\cal B},{\cal N}}{}^{{\cal Q}}\,
\Theta_{{\cal P}}\Theta_{{\cal Q}} ~=~ 0
\;.
\label{Quadratic}
\eea
It has been shown in~\cite{Samtleben:2007an} that an embedding tensor transforming in the ${\bf 45}_{-2/9}$ as
considered above, automatically satisfies the quadratic constraint (\ref{Quadratic}), i.e.\ any choice of $\theta_{kl}$ 
defines a consistent gauging.
We will confirm that statement in this paper by explicit construction. A first indication is the fact that for any choice 
of $\theta_{kl}$,
 the projected generators (\ref{so9}) close into an algebra.

To summarize, the structure of the gauged theory constructed in this paper follows from the general representation
structure of vector fields and embedding tensor components. We will thus consider a deformation of the Lagrangian (\ref{L0SL9}),
whose local gauge algebra is given by the subalgebra of the global $\mathfrak{sl}_9$ spanned by the generators (\ref{so9}).
This is realized by introducing minimal couplings via the covariant derivatives
\bea
{\cal D}_\mu &=& \partial_\mu - g A_\mu{}^{kl}\, \theta_{mk}\,T_{l}{}^m
\;,
\label{minimalso9}
\eea
with the vector fields $A_\mu{}^{kl}=A_\mu{}^{[kl]}$ from (\ref{susyvec}).
In the following we will absorb  the explicit coupling constant $g$ by rescaling $\theta_{mk}$.
In the next section, we will give the details of this construction and show in particular, 
that it is compatible with maximal supersymmetry.


\section{SO(9) supergravity: the Lagrangian}
\label{sec:so9}

Having discussed maximal supergravity in two dimensions in the appropriate $SL(9)$ covariant formulation, we 
proceed to the construction of the gauged theory, i.e.\ in the following we gauge the compact $SO(9)$ 
subgroup of the global symmetry (\ref{sl9}). Just as to avoid any confusion, let us stress that the resulting
theory possesses {\em two} local $SO(9)$ symmetries. The first, which for distinctiveness we shall refer to as  
$SO(9)_K$ is simply related to the $SL(9)/SO(9)$ coset formulation (\ref{coset}) 
and already present in the ungauged theory discussed above. 
It acts linearly on the fermions and
on the scalar fields ${\cal V}$ by {\em right} multiplication
\bea
\delta {\cal V}_k{}^a &=& {\cal V}_k{}^b\,\Omega^{[ba]}
\;.
\label{SO9K}
\eea
On the other hand , the new gauge symmetry $SO(9)_{\rm gauge}$ to be introduced in the following acts by {\em left} multiplication on
the scalars ${\cal V}$ and rotates the 84 scalars $\phi^{klm}$ according to (\ref{sl9}).
Fully covariant derivatives carry the composite connection $Q_\mu^{ab}$ from (\ref{coset}) w.r.t.\ $SO(9)_K$ 
and the connection (\ref{minimalso9}) w.r.t.\ $SO(9)_{\rm gauge}$.

\subsection{General ansatz}

In a first step, we introduce minimal couplings according to (\ref{minimalso9}) in order to 
ensure invariance of the Lagrangian under local gauge transformations generated by (\ref{so9}).
Eventually, we will set $\theta_{kl}\equiv \delta_{kl}$ such that the gauge group corresponds to
the maximal compact subgroup $SO(9)$ of (\ref{sl9}). Throughout the construction it turns out
to be helpful to keep an arbitrary $\theta_{kl}$ for bookkeeping of the different terms. As a byproduct,
the very same construction also yields the theories with $SO(p,9-p)$ and non-semisimple $CSO(p,q,9-p-q)$ gauge groups,
according to the choice of $\theta_{kl}$.

The general ansatz for the gauged Lagrangian then is the following deformation of~(\ref{L0SL9}):
\bea
{\cal L} &=& {\cal L}_{0, {\rm cov}}
- \frac{1}{4} \,\varepsilon^{\mu \nu} {{\cal{F}}_{\mu \nu}}{}^{kl}{}\theta_{lm}\, {Y_{k}}^{m} 
+{\cal L}_{\rm Yuk} + {\cal L}_{\rm pot}
\;,
\label{Lgauged}
\eea
where ${\cal L}_{0, {\rm cov}}$ is obtained by straightforward covariantization of (\ref{L0SL9}) according to
\bea
Q_\mu^{[ab]} + P_\mu^{(ab)} &\rightarrow&
{\cal Q}_{\mu}^{[ab]} +  {\cal P}_{\mu}^{(ab)} ~\equiv~  {\cal{V}}^{-1}{}^{ak}\, 
\left(\partial_{\mu} {{\cal{V}}_{k}}^{b} -  A_{\mu}{}^{lm} \theta_{mk} \, {{\cal{V}}_{l}}^{b} \right)
\;,\nonumber\\[.5ex]
 \partial_{\mu}  \phi^{klm}&\rightarrow&  {\cal D}_{\mu}  \phi^{klm} ~\equiv~ 
\partial_{\mu} \phi^{klm} - 3 \, A_{\mu}{}^{p[k} \theta_{pq} \, \phi^{lm]q} 
\;,\nonumber\\[.5ex]
{\varphi}_{\mu}^{abc} &\rightarrow&
\widetilde{\varphi}_{\mu}^{abc} ~\equiv~ {{ \cal{V}}_{klm}}^{[abc]}\,{\cal D}_{\mu} \phi^{klm}
\;,\nonumber\\[2ex]
D_{\mu} \psi_{\nu}^{I} &\rightarrow&
{\cal D}_{\mu} \psi_{\nu}^{I} ~\equiv~
\partial_{\mu} \psi_{\nu}^{I} 
+ \frac{1}{4} {{\omega_{\mu}}^{\alpha \beta}} \gamma_{\alpha \beta} \, \psi_{\nu}^{I}
+ \frac{1}{4} {\cal Q}_{\mu}^{ab} \, \Gamma_{IJ}^{ab} \, \psi_{\nu}^{J} 
\;,\nonumber\\
 D_{\mu} \chi^{aI} &\rightarrow&
  {\cal D}_{\mu} \chi^{aI}  ~\equiv~
   \partial_{\mu} \chi^{aI} 
 + \frac{1}{4} {{\omega_{\mu}}^{\alpha \beta}} \gamma_{\alpha \beta} \, \chi^{aI}
 + {\cal Q}_{\mu}^{ab} \, \chi^{bI}
+ \frac{1}{4} {\cal Q}_{\mu}^{bc} \, \Gamma_{IJ}^{bc} \, \chi^{aJ} 
\;,
\eea
and thus gauge invariant by construction. Furthermore,
\bea
{\cal F}_{\mu\nu}{}^{kl} &\equiv&
2\partial_{[\mu} A_{\nu]}{}^{kl} + 2 \, \theta_{pq} \, A_{[\mu}{}^{p[k}  A_{\nu]}{}^{l]q}
\;,
\eea
defines the non-abelian field strength of the vectors $A_\mu{}^{kl}=A_\mu{}^{[kl]}$ which in (\ref{Lgauged}) couples 
to an auxiliary field ${Y_{k}}^{l}$.
In anticipation of the final structure we denote this auxiliary field by the same letter as 
the dual scalar potential defined in (\ref{dualY}) above for the ungauged theory.
The general ansatz for the Yukawa-type couplings ${\cal L}_{\rm Yuk}$ in (\ref{Lgauged}) is the collection
of the most general bilinear fermion couplings
\bea
e^{-1}{\cal L}_{\rm Yuk}  &=&
-\frac{1}{2} \, e^{-1} \rho \, \varepsilon^{\mu \nu} 
\left(
\bar{\psi}_{\nu}^{I} \psi_{\mu}^{J} B_{IJ} +
 \bar{\psi}_{\nu}^{I} \gamma^{3} \psi_{\mu}^{J} \tilde{B}_{IJ} 
 -2i  \bar{\psi}_{2}^{I} \gamma_{\nu} \psi_{\mu}^{J} A_{IJ} \right) 
 + i\rho  \, \bar{\psi}_{2}^{I} \gamma^{\mu} \psi_{\mu}^{J} \tilde{A}_{IJ}\nonumber\\
&&{}
+i \rho  \, \bar{\chi}^{aI} \gamma^{\mu} \psi_{\mu}^{J} C^{a}_{IJ} \; - i \rho  \, \bar{\chi}^{aI} \gamma^{3} \gamma^{\mu} \psi_{\mu}^{J} \tilde{C}^{a}_{IJ} \;
+ \rho  \, \bar{\psi}_{2}^{I} \psi_{2}^{J} D_{IJ} \; + \rho  \, \bar{\psi}_{2}^{I} \gamma^{3} \psi_{2}^{J} \tilde{D}_{IJ}\nonumber\\[.5ex]
&&{}
+ \rho  \, \bar{\chi}^{aI} \psi_{2}^{J} E_{IJ}^{a} \; + \rho   \, \bar{\chi}^{aI} \gamma^{3} \psi_{2}^{J} \tilde{E}_{IJ}^{a} \;
+ \rho \, \bar{\chi}^{aI} \chi^{bJ} F_{IJ}^{ab} \; + \rho \, \bar{\chi}^{aI} \gamma^{3} \chi^{bJ} \tilde{F}_{IJ}^{ab}
\;, 
\label{LYuk}
\eea
with tensors $A$, $B$, $C$, $D$, $E$, $F$, depending on the scalar and auxiliary fields to be determined in the following.
Their appearance in (\ref{LYuk}) implies certain symmetry properties such as
\bea
B_{(IJ)}=\tilde{D}_{(IJ)}=0=\tilde{B}_{[IJ]}=D_{[IJ]}
\;,\qquad
F_{IJ}^{ab}=F_{JI}^{ba}\;,\qquad\tilde{F}_{IJ}^{ab}=-\tilde{F}_{JI}^{ba}
\;,
\eea
and
\bea
\Gamma_{IJ}^{a}C_{IK}^{a}=\Gamma_{IJ}^{a} \tilde{C}_{IK}^{a}=0
\;.
\eea
As is standard in gauged supergravity, couplings of the type (\ref{LYuk}) induce a modification of
the fermionic supersymmetry transformation rules (\ref{susy0}) by introduction of the so-called {\em fermion shifts} according to
\bea
\delta_{\epsilon} \psi_{\mu}^{I} &=& 
{\cal D}_{\mu} \epsilon^{I} - \frac{1}{24} \, \rho^{-1/3} \Gamma_{IJ}^{abc} \Big( \gamma^{\nu} \gamma_{\mu}+\frac{1}{3}  \gamma_{\mu} \gamma^{\nu}  \Big) \gamma^{3}\epsilon^{J} \,\widetilde{\varphi}_{\nu}^{abc}
+ i\left(A_{IJ} + \tilde{A}_{IJ} \gamma^{3} \right) \gamma_{\mu} \epsilon^{J}
\;,\nonumber\\
\delta_{\epsilon} \psi_{2}^{I} &=& 
- \frac{i}{2} \, \rho^{-1} \left( \partial_{\mu} \rho \right) \gamma^{3} \gamma^{\mu} \epsilon^{I}
+ \left(B_{IJ}  + \tilde{B}_{IJ} \, \gamma^{3} \right) \epsilon^{J}
\;,\nonumber\\
\delta_{\epsilon} \chi^{a I} &=& 
\frac{i}{2}\,  \Gamma_{IJ}^{b}  \, \gamma^{\mu}\epsilon^{J} {\cal P}_{\mu}^{ab}  
-\frac{i}6\, \rho^{-{1/3}}   \Big( \delta^{ad}\Gamma_{IJ}^{bc}  -  
\frac{1}{6}\, \Gamma_{IJ}^{abcd} \Big)\, \gamma^{3} \gamma^{\mu}\epsilon^{J} \,\widetilde\varphi_{\mu}^{bcd}
+\left( C^{a}_{IJ}  + \tilde{C}^{a}_{IJ} \gamma^{3} \right) \epsilon^{J}
\;,
\nonumber\\
\label{susyshift}
\eea
with the scalar-dependent tensors $A$, $B$, $C$ parametrizing~(\ref{LYuk}). The bosonic supersymmetry
transformation rules from (\ref{susy0}) remain unchanged, the vector fields transform as determined in (\ref{susyvec})
\bea
\delta_{\epsilon}\,A_{\mu}{}^{kl} &=&   \rho^{-2/9}\Big( 
\bar{\psi}_{\mu}^{I} \epsilon^{J} \Gamma_{IJ}^{ab} 
- \frac{2i}{9}\, \bar{\psi}_{2}^{I} \gamma^{3} \gamma_{\mu} \epsilon^{J} \Gamma_{IJ}^{ab} 
 - 2 i  \, \bar{\chi}^{I[a} \gamma_{\mu} \epsilon^{J} \Gamma_{IJ}^{b]}
  \Big)
 {\cal{V}}^{-1}{}_{[ab]}{}^{kl} 
 \nonumber\\
&&{}
 +2 \, \rho^{-5/9} \Big( 
\bar{\psi}_{\mu}^{I} \gamma^{3} \epsilon^{J} \Gamma_{IJ}^{a}
+ \frac{5i}{9}\, \bar{\psi}_{2}^{I} \gamma_{\mu} \epsilon^{J} \Gamma_{IJ}^{a}
 -  i  \bar{\chi}^{aI} \gamma^{3} \gamma_{\mu} \epsilon^{I} 
 \Big) {\cal{V}}^{-1}{}_{[bc]}{}^{kl}\, \varphi^{abc} 
 \;,
 \label{susyA}
\eea
and the transformation of the auxiliary scalar fields $Y_k{}^l$ is described by (\ref{deY})~below.
Finally, ${\cal L}_{\rm pot}$ in (\ref{Lgauged}) describes a scalar potential whose form is most conveniently 
deduced by demanding the absence of terms proportional to $\bar{\psi}^I_\mu\gamma^\mu \epsilon^I$
in the supersymmetry variation of the full Lagrangian (\ref{Lgauged}):
\bea
{\cal L}_{\rm pot} &\equiv& -e V_{\rm pot}
~=~   -\frac{1}{16}\,e \rho \left( 2  \tilde{A}_{IJ}B_{IJ}-2  A_{IJ}\tilde{B}_{IJ} +  C_{IJ}^{a} C_{IJ}^{a} + \tilde{C}_{IJ}^{a} \tilde{C}_{IJ}^{a} \right)
\;,
\label{Lpot}
\eea
in terms of the Yukawa tensors.
The entire Lagrangian (\ref{Lgauged}) with (\ref{LYuk}), (\ref{Lpot}) thus is given as a function of the scalar functions,
$A$, $B$, $C$, $D$, $E$, $F$, which we will determine explicitly in the next section as functions
of the scalar and auxiliary fields.
Before diving into that calculation, let us finish this section by commenting on the supersymmetry
transformations of the vector and auxiliary fields $A_\mu{}^{kl}$ and $Y_k{}^l$, respectively. 
These have been added as new fields in~(\ref{Lgauged}) and do not appear in the ungauged theory (\ref{L0SL9}).

The role of the vector fields has been discussed in section~\ref{sec:vector} above: the underlying affine symmetry structure
predicts to employ for this gauging 36 vector fields $A_\mu{}^{kl}$ in a finite sub-representation of the basic
representation~(\ref{rep_vector}) of $\mathfrak{e}_{9(9)}$\,. We have determined their 
supersymmetry transformation rules in~(\ref{susyvec}), (\ref{susyA}) by demanding closure of the supersymmetry algebra.
As for the auxiliary fields $Y_k{}^l$, their structure is even more restricted. When checking supersymmetry of the gauged
Lagrangian (\ref{Lgauged}), the result of the ungauged theory (\ref{L0SL9}) is modified by the presence of the gauge
connections as a result of which covariant derivatives no longer commute: 
\bea
[{\cal D}_\mu , {\cal D}_\nu] \; \phi^{klm} &=&
- 3 \, \theta_{pq}\, {\cal F}_{\mu\nu}{}^{p[k}  \, \phi^{lm]q}
\;,\nonumber\\
{\cal D}^{\vphantom{a}}_{[\mu}\,{\cal P}^{ab}_{\nu]} &=& 
\frac{1}{2} \, \theta_{kl}  \,{\cal V}_{m(a}\, {\cal V}^{-1}{}_{b)}{}^{k}\,   {\cal F}_{\mu\nu}{}^{lm} 
\;,
\eea
etc.. Variation of ${\cal L}_{0, {\rm cov}}$ thus leads to a number of terms proportional to the field strength
${\cal F}_{\mu\nu}{}^{kl}$ which we collect as
\bea
\delta_\epsilon \,{\cal L}_{0, {\rm cov}} &=& 
\frac14\varepsilon^{\mu \nu} {{\cal{F}}_{\mu \nu}}{}^{kl}{}\theta_{lm}\; {\Xi_{k}}^{m} 
\;,
\label{extraF}
\eea
up to total derivatives, with
\bea
{\Xi_{k}}^{l} &=&
 \bar{\chi}^{aI} \gamma^{3}\varepsilon^{J} \left(- \rho \, {{\cal{V}}_{k}}^{(a}{\cal{V}}^{-1}{}^{b)l} \, \Gamma_{IJ}^{b} 
 + {\frac13 \, \rho^{1/3} \,  {{\cal{V}}_{k}}^{g}  {\cal{V}}^{-1}{}^{jl} \, \delta^{a[d} \Gamma_{IJ}^{efghi]} \, \varphi^{def} \varphi^{hij}}
  \right) 
\nonumber\\   
&&{} + \frac32\, \rho^{2/3} \, \bar{\chi}^{aI}\varepsilon^{J}\,{\cal{V}}^{-1}{}^{gl} {{\cal{V}}_{k}}^{[a} \varphi^{bc]g} \,\Gamma_{IJ}^{bc}
+ \frac13\,\rho^{2/3} \, \bar{\psi}_{2}^{I} \gamma^{3} \varepsilon^{J} \, {\cal{V}}^{-1}{}^{gl} {{\cal{V}}_{k}}^{a} \varphi^{bcg} \,\Gamma_{IJ}^{abc}   \nonumber\\
&&{} + \bar{\psi}_{2}^{I} \varepsilon^{J} \left( { \frac{1}{2}}\rho \, {\cal{V}}^{-1}{}^{al} {{\cal{V}}_{k}}^{b} \,\Gamma_{IJ}^{ab} 
+ \frac1{54} \, \rho^{1/3} \,  {\cal{V}}^{-1}{}^{gl} {{\cal{V}}_{k}}^{d}\,\varphi^{abc} \varphi^{efg} 
\,\Gamma_{IJ}^{abcdef}  \right)  
\;.
\label{Xi}
\eea
The role of the new coupling of the field strength 
to an auxiliary field $Y_k{}^l$ in (\ref{Lgauged}) is precisely to cancel the unwanted contributions
(\ref{extraF}) by imposing
\bea
\delta_\epsilon\,Y_k{}^l &=& {\Xi_{k}}^{l}
\;.
\label{deY}
\eea
Comparing (\ref{Xi}) to (\ref{susyY}), we deduce, that we may identify the auxiliary field $Y_k{}^l$
with the dual scalar potential defined in (\ref{dualY}) for the ungauged theory. Moreover, the 
vector field equations of the full Lagrangian (\ref{Lgauged}) give rise to the duality equations
\bea
- e \varepsilon_{\mu \nu} \theta_{m[k}  {\cal D}^{\nu} Y_{l]}{}^{m} &=&
\rho \,  \theta_{m[k} {{\cal{V}}_{l]a}}    {\cal{V}}^{-1}{}_b{}^m  \left(
 {\cal P}_{\mu}^{ab}- 
\rho^{-2/3}\,    \varphi^{bcd}   \widetilde\varphi_{\mu}^{acd} \right)  
\nonumber\\
&&{}  + \frac1{54} \, e \varepsilon_{\mu \nu} \,   \varepsilon^{abc def ghi}\,\theta_{m[k} 
{{\cal{V}}_{l]}}^{a}  {\cal{V}}^{-1}{}^{jm}\,\varphi^{bcj}  \varphi^{def} \widetilde\varphi^{\nu \, ghi} 
\nonumber\\[1ex]
&&{} +~{\rm fermions}\;,
\label{dualDY}
\eea
which can be understood as the proper covariantization of 
(a projected version of) the duality equations (\ref{dualY}) of the ungauged theory.

To summarize, the supersymmetry transformation rules of the various fields appearing in the
gauged theory are given in (\ref{susy0}), (\ref{susyshift}),
(\ref{susyA}), and (\ref{deY}). The Lagrangian is of the form (\ref{Lgauged}) with (\ref{LYuk}) and (\ref{Lpot}) 
given in terms of the scalar functions, $A$, $B$, $C$, $D$, $E$, $F$, that we will determine explicitly in the next section.

\subsection{Yukawa tensors}

With the general ansatz for the Lagrangian
specified in (\ref{Lgauged}), (\ref{LYuk}), (\ref{Lpot}), we are now in position to 
explicitly check its transformation under supersymmetry 
using  
(\ref{susy0}), (\ref{susyshift}),
(\ref{susyA}), and (\ref{deY}).
As a first result, we obtain a number of linear relations among the Yukawa tensors 
$A$, $B$, $C$, $D$, $E$, $F$ by demanding that all terms linear in space-time 
derivatives cancel in the supersymmetry variation of (\ref{Lgauged}).
E.g.\ vanishing of the terms proportional to 
$\bar{\psi}_\mu^I \gamma^{\mu\nu} \epsilon^J \,\partial_\nu\rho$ and
$\bar{\chi}^{aI} \gamma^{\mu} \epsilon^J \,\partial_\mu\rho$ imposes the relations
\bea
A_{IJ} - \tilde{B}_{IJ} - \rho \frac{\delta \tilde{B}_{IJ}}{\delta \rho} &=&0
\;,\nonumber\\
C^{a}_{IJ} + 2 \rho  \frac{\delta {C}^{a}_{IJ}}{\delta \rho} +  \tilde{E}^{a}_{IJ}&=&0
\;,
\eea
respectively. A more complete list of all such linear relations is 
collected in appendix~\ref{subsec:linear}.
Upon decomposing the Yukawa tensors into their $SO(9)_K$ irreducible parts,
we find that they are uniquely determined by these linear relations in
terms of the scalar fields ${\cal V}_k{}^a$, $\Phi^{klm}$ and the auxiliary fields $Y_k{}^l$\,.
The final result is obtained after some lengthy calculation and reads\footnote{
For `simplicity' of the expressions we have chosen to give the tensors $E^a_{IJ}$ and $F^{ab}_{IJ}$
(and their tilded analogues) in a form which is not yet explicitly projected onto the gamma-traceless part in the
corresponding indices, e.g.\ $\Gamma^a_{IJ}E^a_{JK}\not=0$, etc. Nevertheless, in the Lagrangian (\ref{LYuk})
all these tensors appear only under projection with the (gamma-traceless) fermions $\chi^{aI}$,
i.e.\ eventually only their gamma-traceless parts contribute to the couplings.}
\bea
  A_{IJ} &=&\frac{7}{9}  \, \delta_{IJ}\, b - \frac{5}{9} \,\Gamma_{IJ}^{a}\, b^{a} {-} \frac{1}{9} \, \Gamma_{IJ}^{abcd}\, b^{abcd}  
  \;,\nonumber\\
 \tilde{A}_{IJ} &=& {-} \frac{2}{9} \, \Gamma_{IJ}^{ab}\, b^{ab} {-} \frac{4}{9} \, \Gamma_{IJ}^{abc}\, b^{abc}
  \;,\nonumber\\
 B_{IJ} &=& {-} \Gamma_{IJ}^{ab} \, b^{ab} {-} \Gamma_{IJ}^{abc}\, b^{abc}
  \;,\nonumber\\
\tilde{B}_{IJ} &=& \delta_{IJ} \, b + \Gamma_{IJ}^{a} \, b^{a} {-} \Gamma_{IJ}^{abcd} \, b^{abcd}
  \;,\nonumber\\
C_{IJ}^{a} &=& \frac{8}{9} \, \delta_{IJ}\, b^{a} - \frac{1}{9} \Gamma_{IJ}^{ab}\, b^{b} {-} \frac{20}{9} \, \Gamma_{IJ}^{bcd}\, b^{abcd} {+} \frac{4}{9} \, \Gamma_{IJ}^{abcde}\, b^{bcde}
 + {c}^{ab} \, \Gamma_{IJ}^{b}
  \;,\nonumber\\
\tilde{C}_{IJ}^{a} &=& {+}\frac{14}{9} \,\Gamma_{IJ}^{b}\, b^{ab} {-} \frac{2}{9} \, \Gamma_{IJ}^{abc}\, b^{bc} {-} \frac{2}{3} \, \Gamma_{IJ}^{bc}\, b^{abc} {+} \frac{1}{9} \, \Gamma_{IJ}^{abcd}\, b^{bcd} 
+ {c}^{a,bc} \, \Gamma_{IJ}^{bc}
  \;,\nonumber\\
D_{IJ} &=& \frac{14}{81} \, \delta_{IJ} \, b - \frac{70}{81} \, \Gamma_{IJ}^{a} \, b^{a} {-} \frac{8}{81} \, \Gamma_{IJ}^{abcd} \, b^{abcd} 
  \;,\nonumber\\
\tilde{D}_{IJ} &=& {-} \frac{22}{81} \, \Gamma_{IJ}^{ab} \, b^{ab} {+} \frac{20}{81} \, \Gamma_{IJ}^{abc} \, b^{abc}
  \;,\nonumber\\
E_{IJ}^{a} &=&  {-} \frac{26}{9} \,\Gamma_{IJ}^{b}\, b^{ab} {+} \frac{1}{9} \, \Gamma_{IJ}^{bc}\, b^{abc} -\frac{1}{9} \, {c}^{a,bc} \, \Gamma_{IJ}^{bc}
  \;,\nonumber\\
\tilde{E}_{IJ}^{a} &=& \frac{19}{9}\, \delta_{IJ}\, b^{a} {-}  \frac{28}{9} \, \Gamma_{IJ}^{bcd}\, b^{abcd}  - \frac{5}{9} \, {c}^{ab} \, \Gamma_{IJ}^{b}
  \;,\nonumber\\
  F^{ab}_{IJ}&=& -\frac{1}{18} \, \delta^{ab} \delta_{IJ} \, b + \frac{1}{2} \,  \delta^{ab} \, \Gamma_{IJ}^{c} \, b^{c} 
{+} \frac{1}{2} \, \delta^{ab} \, \Gamma_{IJ}^{cdef} \, b^{cdef}
  -12 \, \Gamma_{IJ}^{cd} \, b^{abcd} - 2 \, {c}^{ab} \, \delta_{IJ}
  \;,\nonumber\\
   \tilde{F}^{ab}_{IJ} &=& {-} \frac{1}{2} \, \delta^{ab} \, \Gamma_{IJ}^{cd} \, b^{cd} {+} \frac{1}{2} \, \delta^{ab} \, \Gamma_{IJ}^{cde} \, b^{cde} {+} 2 \, \delta_{IJ} \, b^{ab}
{+} 2 \, \Gamma_{IJ}^{c} \, b^{abc} 
- 2 \, {c}^{c,ab} \, \Gamma_{IJ}^{c}
\;,
\label{A-F}
\eea
expressed in terms of the $SO(9)_K$ irreducible tensors
\bea
  b &=& \frac{1}{4} \, \rho^{-2/9} \, T
  \;,\nonumber\\
  b^{a} &=&
   - \rho^{-14/9} \,T^{cd} \, \varphi^{abc} {\cal Y}^{bd}
+\frac{1}{144} \, \rho^{-14/9} \, \varepsilon^{bcdefghij} T^{kl} \varphi^{kef} \varphi^{lgh} \varphi^{aij} \varphi^{bcd}
 \;,\nonumber\\
  b^{ab} &=& -\frac{1}{2} \, \rho^{-11/9} \,  T^{d[a} {\cal Y}^{b]d} 
  +\frac{1}{144}\, \rho^{-11/9} \, \varepsilon^{abcdefghi} T^{jk} \varphi^{jcd} \varphi^{kef} \varphi^{ghi}
   \;,\nonumber\\
  b^{abc} &=& \frac{1}{4} \, \rho^{-5/9} \, T^{d[a} \varphi^{bc]d} 
   \;,\nonumber\\
    b^{abcd} &=&- \frac{1}{8} \, \rho^{-8/9} \, T^{ef} \varphi^{e[ab} \varphi^{cd]f}
     \;,\nonumber\\
 {c}^{ab} &=&  - \frac{1}{2} \, \rho^{-2/9} \Big( T^{ab}- \frac{1}{9} \delta^{ab} T \Big) 
  \;,\nonumber\\
 {c}^{a,bc} &=& \frac{1}{3} \, \rho^{-5/9} \left( T^{da} \varphi^{bcd} - T^{d[b} \varphi^{c]ad}  \right) 
 \;,
 \label{bc}
\eea
where we have defined
\bea
T^{ab} &\equiv& {\cal{V}}^{-1}{}^{(kl)}{}_{ab}\,\theta_{kl} \;,\quad T~\equiv~ T^{aa}
\;,\quad
\nonumber\\[.5ex]
\varphi^{abc} &\equiv& {\cal V}_{[klm]}{}^{abc} \phi^{klm}
\;,\quad
{\cal Y}^{ab} ~\equiv~
{\cal V}^{-1}{}^{ak} \,{\cal V}_l{}^b\,Y_k{}^l
\;.
\eea 
It may seem remarkable, that the highly overdetermined system (\ref{rel1})--(\ref{rel4}) of linear relations
for the Yukawa tensors admits a non-trivial solution (\ref{A-F}), (\ref{bc}). With hindsight, this is a further confirmation that the 
algebraic framework which determines the gauge couplings based on the underlying affine 
symmetry~\cite{Samtleben:2007an} is indeed
compatible with supersymmetry.

A further (and final) test to the construction comes from vanishing of the terms that are bilinear in $\theta_{kl}$
in the supersymmetry variation of (\ref{Lgauged}). E.g.\ cancellation of the terms proportional to
$\bar{\psi}^I_\mu \gamma^\mu\gamma^3 \epsilon^J$ implies the relation
\bea
2A_{K(I}B_{J)K}+ 2\tilde A_{K(I}\tilde B_{J)K}+C^a_{KI} \tilde{C}^a_{KJ} +C^a_{KJ} \tilde{C}^a_{KI} &=& 0
\;.
\label{quadAB}
\eea
Employing the linear constraints~(\ref{rel1}) this relation reduces to
\bea
\left( 4 + 2 \rho \partial \rho \right) ( B_ {K(I} \tilde{B}_{J)K} ) =  C^a_{KI} \tilde{C}^a_{KJ} +C^a_{KJ} \tilde{C}^a_{KI} 
\nonumber
\eea
With the explicit parametrization (\ref{A-F})
in terms of $SO(9)_K$ irreducible tensors, the l.h.s.\ and r.h.s.\ of this equation become
\bea
\left( 4 + 2 \rho \partial \rho \right) ( B_ {K(I} \tilde{B}_{J)K} )  &=& 
\Gamma^{b}_{IJ} \left( \frac{80}{3} \, b^{def} b^{defb}-\frac{28}{9} \, b^{a} b^{ab}  \right) 
 + \frac29 \,\Gamma^{aefg}_{IJ} \left( 8 \, b^{ab} b^{befg} -  b^{a} b^{efg} \right) 
\nonumber\\
 &&{}
 - \frac{40}{3} \, \Gamma^{abefg}_{IJ} \, b^{abc} b^{cefg} \;,
 \nonumber\\[1ex]
 C^a_{KI} \tilde{C}^a_{KJ} +C^a_{KJ} \tilde{C}^a_{KI}  &=&
\Gamma^{b}_{IJ} \left(  \frac{80}{3} \, b^{def} b^{defb} -\frac{28}{9} \, b^{a} b^{ab} \right) 
+ \frac29 \,\Gamma^{aefg}_{IJ} \left( 8 \, b^{ab} b^{befg} -  b^{a} b^{efg} \right)
\nonumber\\
 &&{}
+ \Gamma^{abefg}_{IJ} \left(\frac{8}{3} \, b^{abc} b^{cefg} -8 \, c^{c,ab} b^{cefg} \right)
\;,\nonumber
 \eea
respectively. Eventually, the quadratic relation (\ref{quadAB}) thus boils down to the relation
\bea
c^{f,[ab} b^{cde]f} &=& 2 \, b^{f[ab} b^{cde]f}
  \;,
\eea
for the tensors $b^{abc}$, $b^{abcd}$, $c^{a,bc}$. With the explicit form (\ref{bc})
of these tensors, it is straightforward to verify, that this equation is indeed
identically satisfied. 

Supersymmetry imposes several other quadratic conditions analogous to (\ref{quadAB})
on the Yukawa tensors, which we have collected in appendix~\ref{subsec:quadratic}. Just as for (\ref{quadAB}), 
straightforward but lengthy computation shows that all these quadratic equations are identically satisfied 
with the explicit form (\ref{A-F}), (\ref{bc}) 
of the Yukawa tensors.\footnote{Part of these calculations have been facilitated by use of the computer algebra
system Cadabra~\cite{Peeters:2006kp,Peeters:2007wn}.}
This completes the main result of this section: the Lagrangian (\ref{Lgauged}) with (\ref{LYuk}), (\ref{Lpot})
and the Yukawa tensors given in (\ref{A-F}), (\ref{bc}) above
is maximally supersymmetric (up to higher fermion terms and total derivatives).
This result confirms the prediction of~\cite{Samtleben:2007an} discussed in section~\ref{subsec:embedding} above
that any embedding tensor $\theta_{kl}$ defines a consistent gauged theory
compatible with maximal supersymmetry.
In the following, we will set $\theta_{kl}=g\delta_{kl}$ thereby specifying the construction to the $SO(9)$ theory
while making the gauge coupling constant manifest.
Accordingly, we use the $\delta_{kl}$ symbol to raise and lower the corresponding indices.

Let us finally note that with the Yukawa tensors determined in terms of the scalar fields, 
we may evaluate the scalar potential (\ref{Lpot}) of the theory and obtain
\bea
 {V}_{\rm pot}&=& 
 \rho \Big( 2 \,  b^{a} b^{a} + 4 \, b^{ab} b^{ab}  + 48 \, b^{abcd} b^{abcd} 
  + {c}^{ab} {c}^{ab} + 2 \, {c}^{a,bc} \,{c}^{a,bc} 
  - \frac{14}{9} \, bb - 4 \, b^{abc} b^{abc} 
    \Big)
 \;,
 \nonumber\\
 \label{potbc}
\eea
in terms of the $SO(9)_K$ irreducible tensors~(\ref{bc}).

\subsection{Supersymmetry algebra}
\label{subsec:susy_algebra}

Having shown that the Lagrangian (\ref{Lgauged}) defines a maximally supersymmetric theory
with local gauge group $SO(9)$, it is instructive to analyze the structure of the supersymmetry
algebra of this model.
On the bosonic fields appearing in (\ref{L0SL9}), the supersymmetry algebra closes in a 
standard fashion
\bea
[\delta_{\epsilon_1},\delta_{\epsilon_2}] &=&
\xi^\mu\,{\cal D}_\mu + \delta^{\rm Lorentz}_{\omega}+ \delta^{SO(9)_K}_{\Omega}+
\delta^{SO(9)_{\rm gauge}}_\Lambda 
\;,
\label{closure}
\eea
into the local bosonic symmetries of the gauged theory.
As usual,
$\xi^\mu\,{\cal D}_\mu$ denotes a covariant general coordinate transformation
with parameter $\xi^\mu$, combining a spacetime
diffeomorphism with the gauge transformations
of the form
\bea
\omega^{\alpha\beta} = -\xi^\mu\omega_\mu{}^{\alpha\beta} \;,\quad
\Omega^{ab} = -\xi^\mu {\cal Q}_\mu^{ab}\;,\quad
\Lambda^{kl} = -\xi^\mu A_\mu{}^{kl}
\;.
\eea
The r.h.s.\ of the supersymmetry algebra (\ref{closure}) is given by gauge transformations
with parameters
\bea
\xi^{\mu} &=& i \, \bar{\epsilon}_{2}^{I} \gamma^{\mu} \epsilon_{1}^{I}
\;,\nonumber\\
\omega^{\alpha\beta} &=& 
 -2\,\varepsilon^{\alpha\beta} \left( \bar{\epsilon}_{2}^{I} \gamma^{3} \epsilon_{1}^{J} A_{IJ} 
-\bar{\epsilon}_{2}^{I} \epsilon_{1}^{J} \tilde{A}_{IJ} \right)   
\;,\nonumber\\
\Lambda^{kl} &=& -
 \rho^{-5/9} \,  {\cal{V}}^{-1}{}_{[ab]}{}^{kl}\left(
\rho^{1/3}\,\bar{\epsilon}_{2}^{I} \epsilon_1^{J} \Gamma_{IJ}^{ab} \, 
+ 2 \,  \bar{\epsilon}_{2}^{I} \gamma^{3} \epsilon_{1}^{J} \,\Gamma_{IJ}^{c} \, 
  \varphi^{abc}\right)
\;,\nonumber\\
\Omega^{ab} &=& 
-\Lambda^{kl} \, {\cal V}_{kl}{}^{c[a}\,T^{b]c}    
\;.
\label{ddparameters}
\eea
Closure of the supersymmetry algebra requires several of the identities
among the Yukawa tensors that we have collected in appendix~\ref{app:relations},
as well as their explicit form (\ref{A-F}). E.g.\ evaluating successive supersymmetry transformations
on the scalars ${\cal V}_k{}^a$ yields
\bea
[\delta_{\epsilon_1},\delta_{\epsilon_2}]\,  {{\cal{V}}_{m}}^{a} &=&  \xi^{\mu}\, {\cal P}_{\mu}  {{\cal{V}}_{m}}^{a} 
+ \lambda^{(ab)}\,
{{\cal{V}}_{m}}^{b} 
\;,
\label{ddV}
\eea
with 
\bea
\lambda^{(ab)} &\equiv&
2\,  \left(
\bar{\epsilon}_{2}^{I} \gamma^{3} \epsilon_{1}^{J}   \,\tilde{C}_{K(I}^{(a} \Gamma_{J)K}^{b)}
-  \bar{\epsilon}_{2}^{I} \epsilon_{1}^{J}  \,C_{K[I}^{(a} \Gamma_{J]K}^{b)} \right) 
\nonumber\\
&=& 
2 \, \rho^{-5/9}\, \bar{\epsilon}_{2}^{I} \gamma^{3} \epsilon_{1}^{J} \,
\Gamma_{IJ}^{c}  \,T^{d(a} \varphi^{b)cd}
+ \rho^{-2/9}\,  \bar{\epsilon}_{2}^{I} \epsilon_{1}^{J} \, 
\Gamma^{c(a} T^{b)c}
\nonumber\\
&=& 
\Lambda^{ml} \theta_{kl} {\cal V}_m{}^a {\cal V}^{-1\,kb} - \Omega^{ab}
\;,
\eea
such that the second term in (\ref{ddV}) may be rewritten
\bea
\lambda^{(ab)}\,{{\cal{V}}_{m}}^{b} 
&=&
\Lambda^{kl} \theta_{lm} {\cal V}_k{}^a  + {{\cal{V}}_{m}}^{b} \,\Omega^{ba}
\;,
\eea
as a combination of local $SO(9)_{\rm gauge}$ and $SO(9)_K$ transformations
with the parameters of (\ref{ddparameters}).
Moreover, we find that the parameter of gauge transformations $\Lambda^{kl}$ that arises in this commutator
precisely agrees with what
we have found in lowest order in the closure of the supersymmetry algebra on the vector fields in (\ref{lamlam}).
When computing the algebra of the full supersymmetry transformations (\ref{susyshift}) on the vector fields, 
we obtain
additional contributions from the fermion shifts 
\bea
[\delta_{\epsilon_1},\delta_{\epsilon_2}]\,{A_{\mu}}^{kl} &=& 
{\cal D}_{\mu} \Lambda^{kl}
+2 i \rho^{-2/9}  \,{\cal{V}}^{-1}{}^{kl}{}_{[ab]}
\left(\bar{\epsilon}_{2}^{I} \gamma_{\mu} \epsilon_{1}^{J}\, Z^{ab}_{IJ}-
 \bar{\epsilon}_{2}^{I} \gamma^{3} \gamma_{\mu} \epsilon_{1}^{J} \,
\tilde{Z}^{ab}_{IJ} \right)
\;,
\label{ddA}
\eea
which upon using (\ref{rel1}) and the explicit form (\ref{A-F}), (\ref{bc}), 
of the Yukawa tensors combine into
{\small 
\bea
Z^{ab}_{IJ} &=&
2 C_{K(I}^{[a} \Gamma_{J)K}^{b]} 
-{ \Big( A_{K(I} - \frac{2}{9} \tilde{B}_{K(I}\Big)\,\Gamma^{ab}_{J)K}
  - 2 \rho^{-1/3} \varphi^{abc} \Big[\tilde{C}_{(IJ)}^{c}+\Big( \tilde{A}_{K(I}
+ \frac{5}{9} B_{K(I} \Big) \Gamma_{J)K}^{c}   \Big] }
\nonumber\\
&=&
 \Big[
 2 C_{K(I}^{[a} \Gamma_{J)K}^{b]}-
 \Big( \frac{7}{9} +\rho\partial_\rho \Big) \,\tilde{B}_{K(I}\,\Gamma_{J)K}^{ab} 
 \Big]
-   2 \rho^{-1/3} \varphi^{abc} \Big[ 
 \tilde{C}_{(IJ)}^{c} - \Big( \frac49+\rho\partial_\rho\Big) B_{K(I}  \Gamma_{J)K}^{c}
 \Big]
 \nonumber\\[2ex]
 &=& 0\;,
 \eea}
 and
 {\small
 \bea
 \tilde{Z}^{ab}_{IJ} &=&
 2 \tilde{C}_{K(I}^{[a} \Gamma_{J)K}^{b]}+
 {\Big( \tilde{A}_{K(I} + \frac{2}{9} B_{K(I} \Big) \Gamma_{J)K}^{ab} 
-  2\rho^{-1/3} \varphi^{abc}\Big[C_{(IJ)}^{c} - \Big( A_{K(I} - \frac{5}{9} \tilde{B}_{K(I} \Big) \Gamma_{J)K}^{c}  \Big]}
\nonumber\\
&=&
\Big[2 \tilde{C}_{K(I}^{[a} \Gamma_{J)K}^{b]}-
  \Big( \frac{7}{9} +\rho\partial_\rho \Big) {B}_{K(I}  \Gamma_{J)K}^{ab} \Big]
-  2\rho^{-1/3} \varphi^{abc}\Big[C_{(IJ)}^{c} -
   \Big( \frac49+\rho\partial_\rho\Big) \tilde{B}_{K(I} \Gamma_{J)K}^{c}  \Big]
\nonumber\\
&=& 
4 \,\delta_{IJ} \,
 \rho^{-2/9}  \left( b^{ab} - \rho^{-1/3} \, \varphi^{abc} b^{c} \right)  
  \;.
\eea}
As a result, the commutator on the vector fields (\ref{ddA}) closes into the standard form 
\bea
[\delta_{\epsilon_1},\delta_{\epsilon_2}]\,{A_{\mu}}^{kl} &=& 
{\cal D}_{\mu} \Lambda^{kl}+
\xi^\nu\,  {\cal F}_{\nu\mu}{}^{kl}
\;,
\label{ddAF}
\eea
provided their field strengths satisfy the relation%
\bea
{\cal V}_{kl}{}^{ab}\, {\cal F}_{\mu\nu}{}^{kl} &=& 
 8 \, e  \varepsilon_{\mu \nu} \,\rho^{-2/9} \left( b^{ab} - \rho^{-1/3} \, \varphi^{abc} b^{c} \right) 
 +{\rm fermions}
 \;.
 \label{eomF}
\eea
This in turn are precisely the equations of motion 
obtained by varying the Lagrangian (\ref{Lgauged}) with respect to the
auxiliary field $Y_k{}^l$, using that the corresponding derivative of the scalar potential (\ref{potbc})
takes the form\footnote{
To be precise, let us note that for degenerate choice of $\theta_{kl}$, only a subset 
$\theta_{mk} A_\mu{}^{kl}$ of the vector fields $A_\mu{}^{kl}$
appears in the Lagrangian and consistently (\ref{ddAF}) and (\ref{eomF}) only hold for this subset.
On the other hand, for the $SO(9)$ theory these equations consistently apply to all the vector fields.
}
\bea
\frac{\partial {\cal L}_{\rm pot}}{\partial Y_{kl}} &=&
-4\rho^{-2/9}   \,{\cal{V}}^{-1}{}^{kl}{}_{ab}\,
\left( b^{ab}-
\rho^{-1/3} \varphi^{abc}\,b^c \right)
\;.
\eea
We have thus shown that the supersymmetry algebra of the gauged theory
consistently closes on-shell on the vector fields.
As a final exercise, one may verify, that the supersymmetry algebra also 
closes on-shell on the auxiliary scalar fields $Y_{kl}$
\bea
[\delta_{\epsilon_1},\delta_{\epsilon_2}]\, Y_{kl} &=&
\xi^\mu {\cal D}_{\mu} Y_{kl}
+2\Lambda_{[k}{}^{n}\, Y_{l]n}
\;,
\eea
provided they satisfy the first-order field equations~(\ref{dualDY})
obtained from the Lagrangian (\ref{Lgauged}).\footnote{For degenerate choice of $\theta_{kl}$,
the same restriction discussed in the footnote above applies.}
This yields yet another check for the supersymmetry transformations of these fields
proposed in~(\ref{deY}).


\section{SO(9) supergravity: properties}
\label{sec:properties}

In this paper, we have constructed maximal supergravity in two dimensions
with gauge group $SO(9)$. The resulting theory is described by the Lagrangian (\ref{Lgauged}), with the
different terms defined in (\ref{L0SL9}), (\ref{LYuk}), and (\ref{Lpot}) , respectively.
The Yukawa tensors are explicitly given in (\ref{A-F}), (\ref{bc}) as functions
of the scalar fields.
In this section, we discuss some of the properties of the theory. In particular, we derive the
full set of bosonic field equations and show that the theory admits a domain wall solution that 
preserves half of the supersymmetries.
Finally, we briefly
discuss alternative off-shell formulations of the theory obtained by integrating out
some of the auxiliary fields. In particular, these may be convenient for applications of the theory in the holographic context.

\subsection{The bosonic field equations}

In this section we derive the bosonic equations of motion of the theory.
Variation of the Lagrangian (\ref{Lgauged}) w.r.t.\ the dilaton and the metric gives rise to the equations
\bea
\frac{1}{4} R &=&
 \frac{1}{4}   \, {\cal P}^{\mu \,ab } {\cal P}_{\mu}^{ab}
+ \frac1{36}\, \rho^{-2/3}  \, \tilde\varphi^{\mu \,abc} \tilde\varphi_{\mu}^{ abc} -\frac{\partial V_{\rm pot}}{\partial \rho} 
\;,
\nonumber\\
\nabla^2 \rho
&=&
4 V_{\rm pot}
\;,
\nonumber\\
0&=& \nabla_\mu \partial_\nu \rho 
+ \rho \,{\cal P}_\mu^{ab} {\cal P}_\nu^{ab}
+\frac1{3} \,\rho^{1/3} \tilde\varphi_\mu^{abc} \tilde\varphi_\nu^{abc} 
- \frac12 \,g_{\mu\nu}\,({\rm trace})
\;.
\label{eom:md}
\eea
The first two equations constitute the second order field equations for the conformal factor of the two-dimensional
metric and the dilaton, respectively.
The last (constraint) equation in (\ref{eom:md}) corresponds to variation of the Lagrangian
w.r.t.\ the two unimodular degrees of freedom of the metric, that appear as Lagrange multipliers
as usual in two dimensions.
For the physical scalar fields we obtain the equations
\bea
{\cal D}^\mu \left(\rho\, {\cal P}_\mu^{ab} \right) &=& 
\left({\cal V}_{kl}{}^{ab}-\ft19 \delta^{ab}{\cal M}_{kl}\right)
{\cal M}_{mn} {\cal M}_{pq}\,
{\cal D}^\mu \phi^{kmp} {\cal D}_\mu \phi^{lnq}
-2\,\frac{\partial V_{\rm pot}}{\partial \Sigma^{ab}} \;,
\nonumber\\
{\cal D}_\mu {\cal D}^\mu \left( {\cal N}_{klm,pqr}\, \phi^{pqr} \right)
&=& \frac1{36}\, e^{-1} \varepsilon^{\mu \nu} \varepsilon_{klmnpqrst}\left(
{\cal D}_{\mu} \phi^{npq} \, {\cal D}_{\nu} \phi^{rst}  -  {\cal F}_{\mu\nu}{\,}_u{}^r\,\phi^{npq} \phi^{stu} \right)
\nonumber\\
&&{}
-6\,\frac{\partial V_{\rm pot}}{\partial \phi^{klm}}
 \;.
 \label{scalars}
\eea
with ${\cal M}_{kl}\equiv {\cal V}_k{}^a {\cal V}_l{}^a$, 
${\cal N}_{klm,pqr}\equiv\rho^{1/3}\, {\cal V}_{(klm)}{}^{abc}{\cal V}_{(pqr)}{}^{abc}$,
and the covariant variation scalar defined by
$\delta_\Sigma {\cal V}_m{}^a \equiv {\cal V}_m{}^c \Sigma^{ac}$
with symmetric traceless $\Sigma^{ab}$\,.
Finally, the vector fields and the auxiliary scalars satisfy the first-order equations
\bea
{\cal V}_{kl}{}^{ab}\, {\cal F}_{\mu\nu}{}^{kl} &=& 
 8 \, e  \varepsilon_{\mu \nu} \,\rho^{-2/9} \left( b^{ab} - \rho^{-1/3} \, \varphi^{abc} b^{c} \right) 
 \;,\label{1order}\\[1ex]
  \rho \,  {\cal W}_{kl}{}^{ab}   \left(
 {\cal P}_{\mu}^{ab}- 
\rho^{-2/3}\,    \varphi^{bcd}   \widetilde\varphi_{\mu}^{acd} \right)  
&=&
e \varepsilon_{\mu \nu} \Big({\cal D}^{\nu} Y_{[kl]}
 - \frac1{54}   \varepsilon^{abc def ghi}\,
{\cal W}_{kl}{}^{aj} \,\varphi^{bcj}  \varphi^{def} \widetilde\varphi^{\nu \, ghi} \Big)
\;,
\nonumber
\eea
with the scalar tensor ${\cal W}_{kl}{}^{ab} = \delta_{m[k} {{\cal{V}}_{l]}{}^{a}}    {\cal{V}}^{-1}{}^{b m}$\,.
It is straightforward to observe that for the $SO(9)$ theory with $\theta_{kl}=\delta_{kl}$, only the antisymmetric
components $Y_{[kl]}$ of the auxiliary scalar fields enter the Lagrangian such that we may simply omit the symmetric
part $Y_{(kl)}$.
While the first of the first-order equations (\ref{1order}) 
does not impose any integrability relations (the Bianchi identities
in two dimensions are trivial), one may wonder about the consistency of the second 
non-abelian duality equation.
Contracting both sides of this equation with a derivative ${\cal D}^\mu$, and
using the equations of motion (\ref{scalars}), 
leads to
\bea
  -2\,{\cal W}_{kl}{}^{ab} \,\frac{\partial V_{\rm pot}}{\partial \Sigma^{ab}}  
  -6 \,\frac{\partial V_{\rm pot}}{\partial \phi^{mn[k}}\,\phi^{\vphantom{l}}_{l]}{}^{mn}
&=& 
 e^{-1} \varepsilon^{\mu \nu} {\cal F}_{\mu\nu}{}^{m}{}_{[k} Y_{l]m}
\nonumber\\
&=&
-4\,\frac{\delta V_{\rm pot}}{\delta Y^{m[k}}\, Y^{\vphantom{l}}_{l]}{}^m
\;,
\eea
where we have used the first equation of~(\ref{1order}) in the second equality.
Together, we find that consistency of the non-abelian duality equations (\ref{1order})
precisely translates into $SO(9)$ gauge invariance of the scalar potential $V_{\rm pot}$, 
which is guaranteed by construction of (\ref{Lpot}). We thus have a consistent set of 
first- and second-order bosonic field equations.

Finally, it is instructive to give the scalar potential (\ref{Lpot}), (\ref{potbc}), in a more explicit form.
With the explicit expressions (\ref{bc}) for the Yukawa tensors, this potential becomes an eighth order polynomial
in the scalars $\phi^{klm}$, which when expanded to quadratic order takes the form 
\bea
 V_{\rm pot}
 &=&   \frac{1}{8} \, \rho^{5/9} \left( 2 \, {\rm tr}[{\cal{M}}^{-1}{\cal{M}}^{-1}]  - \left({\rm tr}[{\cal{M}}^{-1}]\right)^2   
 \right)
 + \frac{1}{4} \, \rho^{-1/9}  {\cal{M}}_{mp} {\cal{M}}_{nq} {\cal{M}}^{-1}{}_{kl}\, \phi^{mnk} \phi^{pql} 
\nonumber\\[.5ex]
&&{}+ \rho^{-13/9} \left({\cal{M}}^{-1 \, km} {\cal{M}}^{-1 \, ln}  
+ 2 \, \rho^{-2/3} \phi^{klp} {\cal{M}}_{pq}  \phi^{qmn}\right)  Y_{kl}  Y_{mn}
~+{\cal O}(\phi^3)
\;.
\label{pot_exp}
\eea
The first term corresponds to the standard potential of a sphere reduction, see e.g.~\cite{Cvetic:2000dm}.
The dilaton prefactor is a sign of the warped geometry of the reduction. Its presence implies that the 
two-dimensional theory supports a domain wall solution which we will discuss in the next section.

\subsection{Domain wall solution}

From its higher-dimensional origin, we expect the $SO(9)$ theory to describe the fluctuations
around a warped $AdS_2 \times S^8$ geometry. The warping of the higher-dimensional geometry translates into
the fact that the ground state of the lower-dimensional theory is not a pure $AdS$ geometry but rather
a half-supersymmetric domain wall solution~\cite{Boonstra:1998mp,Bergshoeff:2004nq,Bergshoeff:2012pm}.
In order to identify this ground state of (\ref{Lgauged}), we consider the
Killing spinor equations of the theory given by imposing vanishing of the fermionic supersymmetry 
transformations~(\ref{susyshift}).
Let us evaluate these equations at the origin of the scalar target space,
i.e.\ for ${\cal V}_m{}^a=\delta_m^a$ and $\phi^{klm}=0=Y_{kl}$.
In this truncation, the Killing spinor equations reduce to
\bea
 0 &\stackrel{!}=& \delta_\epsilon \psi^I_\mu ~=~ {\cal D}_{\mu} \epsilon^{I} +  \frac{7i}{4}\, g\,\rho^{-2/9} \gamma_{\mu} \epsilon^{I}  \;,\nonumber\\
0&\stackrel{!}=& \delta_\epsilon \psi^I_2 ~=~  - \frac{i}{2} \rho^{-1} \left(  \partial_{\mu} \rho \right) \gamma^{3} \gamma^{\mu} \epsilon^{I} 
+ \frac{9}{4} \,g\,\rho^{-2/9} \gamma^{3} \epsilon^{I}  \;,
\label{KS}
\eea
while $\delta_\epsilon \chi^{aI}=0$ is automatically satisfied.
With the standard domain wall ansatz for the metric
\bea
ds^{2} &=& e^{2A(r)} \, dt^{2} - dr^{2}
\;,
\eea
and assuming a Killing spinor of the form $\epsilon^{I} = f(r) \, \epsilon_{0}^{I}$,
equations (\ref{KS}) are solved by
\bea
f(r)&=& f_0\, r^{7/4}\;,\qquad
A(r)=A_0 + \frac72\, \ln r \;,\qquad
\rho(r)= (g r)^{9/2}
\;,
\label{dw}
\eea
with a constant spinor $\epsilon_0^I$ satisfying the projection condition
\bea
\gamma^{1} \epsilon_{0}^{I} = -i\, \epsilon_{0}^{I} 
\;,
\label{half_susy}
\eea
which breaks supersymmetry to 1/2.
The Ricci scalar for this metric becomes
\begin{equation}
R= \frac{35}{2} \frac{1}{r^{2}}
\;,
\end{equation}
and it is straightforward to show that the solution (\ref{dw}) satisfies the bosonic equations of motion~(\ref{eom:md}).
This is the two-dimensional domain wall solution corresponding to the D0-brane near-horizon 
geometry~\cite{Boonstra:1998mp,Bergshoeff:2004nq}.

\subsection{Auxiliary fields}

We have constructed the $SO(9)$ theory in a form (\ref{Lgauged}) that carries two types of
auxiliary fields: the scalar fields $Y_k{}^l$ and the vector fields $A_\mu{}^{kl}$.
The latter may be integrated out as in~\cite{delaOssa:1992vc} giving rise to yet another
T-duality transformation of the scalar target space such that the resulting theory will
be described by an {\em ungauged} (dilaton-gravity coupled)
non-linear $\sigma$-model on a yet different 128-dimensional target space with Wess-Zumino term.
In this frame, no vector fields are present, and the only remnants of the gauging are the Yukawa couplings and
the scalar potential which is still given by (\ref{pot_exp}).

A more interesting alternative presentation of the theory is obtained by integrating out
the auxiliary scalar fields $Y_{kl}$ which can be expressed in terms of the
non-abelian field strength by equation~(\ref{1order}), explicitly given by
\bea
 {\cal F}_{\mu\nu}{}^{kl} &=& 
 4g  e \, \varepsilon_{\mu \nu} \,\rho^{-13/9} \left(
 {\cal M}^{-1\,kp} {\cal M}^{-1\,lq} 
   +  2 \rho^{-2/3} \, \phi^{klm} {\cal M}_{mn} \phi^{npq} \right) Y_{pq}  
\nonumber\\
&&{}
+\frac{ge}{18} \,   \varepsilon_{\mu \nu} \,\rho^{-13/9}   
 \left({\cal M}^{-1\,kp} {\cal M}^{-1\,lq} 
 +\rho^{-2/3} \, \phi^{klm}  {\cal M}_{mn}  \phi^{npq}
 \right) 
\varepsilon_{pqrstuvxy}  
\, \phi^{zrs} \phi_z{}^{tu} \phi^{vxy} 
\;,
\nonumber\\[.5ex]
&&{}+ {\rm fermions}
 \label{FY}
\eea
upon inversion of the matrix
\bea
{\cal O}^{kl,pq} &\equiv&
{\cal M}^{-1\,kp} {\cal M}^{-1\,lq} 
   +  2 \,\rho^{-2/3} \, \phi^{klm} {\cal M}_{mn} \phi^{npq}
   \;,
\eea
with ${\cal M}_{kl}\equiv {\cal V}_k{}^a {\cal V}_l{}^a$\,.
After integrating out the scalar fields $Y_{kl}$, the bosonic sector of the theory is
described by 128 physical scalars coupled to dilaton gravity together with 36 vector fields.
The latter arise with a two-dimensional Yang-Mills term of the form
\bea
{\cal L}_{{\cal F}^2} &\propto& 
 e \rho^{13/9}\,{\cal F}_{\mu\nu}{}^{kl}\, {\cal O}^{-1}_{kl,mn}\,{\cal F}^{\mu\nu\,mn}
\;,
\label{YM}
\eea
which is of the form as what should follow from the warped sphere reduction.
This is the formulation of the theory that will be most useful in order to
understand its embedding into higher dimensions and to address issues of holography.
We note that  this Lagrangian allows
for a smooth limit $g\rightarrow0$. 
In this limit e.g.\ the kinetic term (\ref{YM}) reduces
to the corresponding coupling of abelian Maxwell field strengths and the Yukawa couplings
formerly carrying $Y_{kl}$ give rise to non-vanishing Pauli-type couplings.
In fact, 
this limit is nothing but the original ungauged theory (\ref{Lag_red}) 
obtained by dimensional reduction 
on the torus  in which the Kaluza-Klein field strengths $F_{\mu\nu\,k}$ have been 
eliminated by use of (\ref{FTheta}) with $\theta^k=0$
(which precisely gives rise to the operator ${\cal O}^{-1}$).


\section{Conclusions}


In this paper, we have constructed maximal supergravity in two dimensions with gauge group~$SO(9)$.
The starting point has been the proper embedding (\ref{soso}) of the gauge group into
infinite-dimensional symmetry group $E_{9(9)}$ of the ungauged theory.
Accordingly, we have performed the construction in a scalar frame in which the $SO(9)$ gauge group
is part of the off-shell symmetries of the Lagrangian. In this frame, the bosonic part of the Lagrangian is 
given by a (dilaton-)gravity coupled 
non-linear gauged $\sigma$-model on a 
target space $\left(SL(9)\ltimes {\mathbb{T}}_{84}\right)/SO(9)$ with Wess-Zumino term. 
We have given the explicit Lagrangian for the gauged theory,
including the expressions for the fermionic sector, the Yukawa couplings and the scalar potential.
\smallskip

To our knowledge, this is the first complete example of a non-trivial gauging of the maximal theory
in two dimensions (apart from the obvious candidates obtained by torus reduction of
higher-dimensional gauged supergravities). It exhibits several characteristic features of the two-dimensional
gaugings such as the appearance of auxiliary scalar fields (here the $Y_{kl}$)
that can be identified within the infinite tower of dual scalar fields of the ungauged theory.
Of course it would be highly interesting to extend this construction to the general gauging
of two-dimensional supergravity, completing the bosonic construction of~\cite{Samtleben:2007an}
by Yukawa couplings and the general scalar potential. Among the technical challenge of this
construction is the decomposition of the embedding tensor in
the basic representation of the affine $E_{9(9)}$ under its compact subgroup 
$K(E_9)$, under which the fermions transform, cf.~\cite{Nicolai:2004nv,Paulot:2006zp}.
From this more general point of view, another highly interesting direction to pursue is the question
to which extent similar constructions can be achieved upon further dimensional reduction.
For one-dimensional supergravity, the structures of the embedding tensor and its supersymmetric
couplings should find their place in the realm of the hyperbolic algebra $E_{10}$, c.f.~\cite{Damour:2006xu}.
\smallskip

The concrete model we have constructed in this paper on the other hand may serve as 
an advanced tool in the study of non-conformal holographic 
dualities~\cite{Itzhaki:1998dd,Boonstra:1998mp,Behrndt:1999mk},
as discussed in the introduction.
The theory dual to $SO(9)$ supergravity is the super matrix quantum mechanics,
obtained by dimensional reduction of ten-dimensional SYM theory
to one dimension, where it is of the form
\bea
{\cal L} &=& {\rm tr}\left\{
(D_t X^k)^2 + \psi^I D_t \psi^I
- \frac12[X_k,X_l]^2 - \Gamma^k_{IJ}\, \psi^I [X_k, \psi^J] 
\right\}
\;,
\label{sym}
\eea
with $SU(N)$ valued matrices $X_k$, $\psi^I$ in the corresponding representation of $SO(9)$. This model
itself has been proposed as a non-perturbative definition of
M-theory~\cite{Banks:1996vh}. 
Some aspects of this correspondence
have been tested, see e.g.~\cite{Youm:1999vn,Anagnostopoulos:2007fw,Catterall:2008yz,Hanada:2011fq},
on the supergravity side however mainly restricted to the dilaton-gravity sector.
With the theory constructed in this paper we have extended the non-propagating dilaton-gravity sector
to include the full non-linear scalar and fermionic couplings of the lowest $N=16$ matter multiplet.
This may allow for various new
tests/applications of the correspondence, in particular involving higher-dimensional
gauge invariant operators on the SYM side (\ref{sym}).
The techniques developed in~\cite{Wiseman:2008qa,Kanitscheider:2008kd} will play
a central role in such an analysis.
Another interesting topic in this context is the explicit higher-dimensional embedding of our model 
and the issue of consistent truncation of warped reductions~\cite{Cvetic:2000yp} for the $S^8$ sphere.
\smallskip

We hope to come back to some of these issues in the future.

\bigskip

\subsection*{Acknowledgments}

It is a pleasure to thank A.~Kleinschmidt, T.~Nutma, M.~Trigiante, and D.~Tsimpis
for very useful discussions.
%

\bigskip
\bigskip

\section*{Appendix}

\begin{appendix}

\section{Relations among Yukawa tensors}
\label{app:relations}

Supersymmetry of the Lagrangian (\ref{Lgauged}) requires
a number of linear, differential, and quadratic relations among the Yukawa tensors 
$A$, $B$, $C$, $D$, $E$, $F$ introduced in (\ref{LYuk}). 
In this appendix we list these relations, ordered by their origin.
They have been used in the main text in order to 
find the (unique) solution (\ref{A-F}), (\ref{bc})
for the Yukawa tensors in terms of the scalar fields.

\subsection{Linear relations among the Yukawa tensors}
\label{subsec:linear}

Demanding that all terms linear in space-time 
derivatives cancel in the supersymmetry variation of (\ref{Lgauged})
implies a number of relations linear in the Yukawa tensors.
The cancellation of terms carrying $\partial_\mu\rho$ induces
\begin{align}
\label{rel1}
A_{IJ} - A_{JI}&=0
\;,&
\tilde{A}_{IJ} + \tilde{A}_{JI}&=0
\;,\nonumber\\
A_{IJ} - \tilde{B}_{IJ} - \rho \frac{\partial \tilde{B}_{IJ}}{\partial \rho} &=0
\;,&
\tilde{A}_{IJ} + B_{IJ} + \rho \frac{\partial B_{IJ}}{\partial \rho} &=0
\;,\nonumber\\
{D}_{IJ} + \rho \frac{\partial {A}_{IJ}}{\partial \rho} &= 0
\;,&
\tilde{D}_{IJ} + \rho \frac{\partial \tilde{A}_{IJ}}{\partial \rho} &= 0
\;,\nonumber\\
C^{a}_{IJ} + 2 \rho  \frac{\partial {C}^{a}_{IJ}}{\partial \rho} +  \tilde{E}^{a}_{IJ}&=0
\;,&
\tilde{C}^{a}_{IJ} + 2 \rho  \frac{\partial {\tilde{C}}^{a}_{IJ}}{\partial \rho} -  E^{a}_{IJ}&=0
\;.
\end{align}
The cancellation  of terms carrying $\widetilde\varphi_\mu^{abc}$ induces
\bea
0&=&
3 \tilde{C}^{[a}_{K[I} \Gamma^{bc]}_{J]K} + \frac{1}{3} B_{K[I} \Gamma_{J]K}^{abc} 
- 3 \, \rho^{-5/9} T^{de} \Gamma_{IJ}^{d[a} \varphi^{bc]e} 
\;,\\
0&=&
3 C^{[a}_{K(I} \Gamma^{bc]}_{J)K} + \frac{1}{3} \, \tilde{B}_{K(I} \Gamma_{J)K}^{abc}
+ 6 \, \rho^{-8/9} \Gamma_{IJ}^{d} T^{ef} \varphi^{de[a} \varphi^{bc]f}\;,\nonumber\\
0&=&
\rho^{1/3} \, \frac{\partial \tilde{B}_{IJ}}{\partial \varphi^{abc}} 
+ \frac{1}{9} \, {B}_{K(I} \Gamma^{abc}_{J)K} - \frac{1}{2} \, \tilde{C}^{[a}_{K(I} \Gamma^{bc]}_{J)K} -\frac{1}{54} \, \rho^{-11/9} \Gamma_{IJ}^{d} \varepsilon^{efghijabc} T^{kl} \varphi^{dke} \varphi^{lfg} \varphi^{hij} \;,
\nonumber\\
0&=&
\rho^{1/3} \, \frac{\partial B_{IJ}}{\partial \varphi^{abc}} - \frac{1}{9} \, \tilde{B}_{K[I} \Gamma^{abc}_{J]K} + \frac{1}{2} \, C^{[a}_{K[I} \Gamma^{bc]}_{J]K}  \nonumber
-\frac{2}{27} \, \rho^{-8/9} T^{i[d} \Gamma_{IJ}^{efghabc]}  \varphi^{ide} \varphi^{fgh} .
\eea\\
The cancellation  of terms carrying ${\cal P}_\mu^{ab}$ induces
\begin{align}
&0= 2 C_{K[I}^{(a} \Gamma_{J]K}^{b)} + \rho^{-2/9} \Gamma_{IJ}^{c(a} T^{b)c} 
\;,&
0&={\cal{P}}_{\mu}^{ab} \Big( \frac{\partial B_{IJ}}{\partial {\Sigma^{ab}}} +3 \, \varphi^{acd}   \frac{\partial B_{IJ}}{\partial \varphi^{bcd}} 
+  \tilde{C}^{a}_{K[I} \Gamma^{b}_{J]K} \Big)
\;,\nonumber\\
&0=2\tilde{C}_{K(I}^{(a} \Gamma_{J)K}^{b)} - 2 \, \rho^{-5/9} \Gamma_{IJ}^{c} T^{d(a} \varphi^{b)cd}
\;,&
0&={\cal{P}}_{\mu}^{ab} \Big(  \frac{\partial \tilde{B}_{IJ}}{\partial {\Sigma^{ab}}} + 3 \, \varphi^{acd}   \frac{\partial \tilde{B}_{IJ}}{\partial \varphi^{bcd}} 
- {C}^{a}_{K(I} \Gamma^{b}_{J)K} \Big)
\;,\nonumber\\[.5ex]
&0=\makebox[0in][l]{
$\displaystyle{\chi^{aJ} {\cal{P}}_{\mu}^{eb} \,\Big( \frac{\partial {C}_{JI}^{a}}{\partial {\Sigma^{eb}}}
 +3  \varphi^{cde}   \frac{\partial {C}^{a}_{JI}}{\partial \varphi^{bcd}} 
- \frac{1}{2} \delta^{ea} \tilde{B}_{KI} \Gamma^{b}_{JK} -  \Gamma^{e}_{KI} F^{ab}_{JK}
- \rho^{-2/9} \Gamma_{IJ}^{c} T^{b[c} \delta^{a]e} \Big)}$\;,}
\nonumber\\
&0= \makebox[0in][l]{
$\displaystyle{\chi^{aJ} {\cal{P}}_{\mu}^{eb} \,\Big(\frac{\partial \tilde{C}_{JI}^{a}}{\partial {\Sigma^{eb}}} + 3 \varphi^{cde}   
\frac{\partial \tilde{C}^{a}_{JI}}{\partial \varphi^{bcd}} 
- \frac{1}{2} \delta^{ea} {B}_{KI} \Gamma^{b}_{JK} +   \Gamma^{e}_{KI} \tilde{F}^{ab}_{JK}
+ \rho^{-5/9} \delta_{IJ} T^{ce} \varphi^{abc} \Big)}$\;,}
\nonumber\\
\end{align}
with the $SO(9)$ covariant variation $\partial/\partial \Sigma^{ab}$ introduced in equation (\ref{scalars}).
The cancellation  of terms carrying ${\cal D}_\mu Y_k{}^l$ finally induces
 \begin{align}
 \frac{\partial A_{IJ}}{\partial {Y_{k}}^{l}} -
\frac{5}{9}  \, \rho^{-14/9} \, \Gamma_{IJ}^{a}   {\theta}_{ml}\, {\cal{V}}^{-1}{}^{km}{}_{bc}  \, \varphi^{abc} &=0
\;,&
 \frac{\partial \tilde{A}_{IJ}}{\partial {Y_{k}}^{l}} +
\frac{1}{9}  \rho^{-11/9} \, \Gamma_{IJ}^{ab}  {\theta}_{ml}\, {\cal{V}}^{-1}{}^{km}{}_{ab} &=0
\;,\nonumber\\[1ex]
\frac{\partial B_{IJ}}{\partial {Y_{k}}^{l}} + 
\frac{1}{2} \, \rho^{-11/9} \,\Gamma_{IJ}^{ab} {\theta}_{ml} \,{\cal{V}}^{-1}{}^{km}{}_{ab}  &=0  
\;,&
\frac{\partial \tilde{B}_{IJ}}{\partial {Y_{k}}^{l}} + 
 \, \rho^{-14/9} \Gamma_{IJ}^{a} {\theta}_{ml}\, {\cal{V}}^{-1}{}^{km}{}_{bc}  \, \varphi^{abc} &=0
\;,\nonumber\\[1ex]
   \frac{\partial C_{IJ}^{a} }{\partial {Y_{k}}^{l}} +
 \, \rho^{-14/9} \, \delta_{IJ}  {\theta}_{ml}\, {\cal{V}}^{-1}{}^{km}{}_{bc}  \, \varphi^{abc} &=0
\;,&
 \frac{\partial \tilde{C}_{IJ}^{a} }{\partial {Y_{k}}^{l}} -
  \rho^{-11/9} \, \Gamma_{IJ}^{b}  {\theta}_{ml}\, {\cal{V}}^{-1}{}^{km}{}_{[ab]}&=0
  \;.
    \label{rel4}
\end{align}

\subsection{Quadratic relations among the Yukawa tensors}
\label{subsec:quadratic}

The remaining identities that supersymmetry imposes on the Yukawa tensors
are bilinear in these tensors. They lead to the following set of equations
\bea
0 &=& 2A_{K(I}B_{J)K}+ 2\tilde A_{K(I}\tilde B_{J)K}+C^k_{KI} \tilde{C}^k_{KJ} +C^k_{KJ} \tilde{C}^k_{KI}
\;,
\nonumber\\
 0 &=&   2 \tilde{B}_{K(I} A_{J)K} + 2 B_{K(I} \tilde{A}_{J)K} - C_{KI}^{a}  C_{KJ}^{a} - \tilde{C}_{KI}^{a} \tilde{C}_{KJ}^{a}    
 + \frac{1}{2} \, \rho^{-1}\,\delta_{IJ} \, \frac{\partial {V}_{\mathrm{pot}}}{\partial \sigma}  \;,
\nonumber\\[1ex]
  0 &=& - 4  A_{K(I} \tilde{A}_{J)K} + 2 D_{IK} B_{KJ} + 2  \tilde{D}_{IK} \tilde{B}_{KJ} + E_{KI}^{a} C_{KJ}^{a} - \tilde{E}_{KI}^{a} \tilde{C}_{KJ}^{a}  
 \nonumber\\[.5ex]
 &&{}
-   \frac{\partial {V}_{\mathrm{pot}} }{\partial {{Y_{k}}^{l}}} \, \Big( {\frac{1}{2}}  \, {\cal{V}}^{-1}{}^{al} {{\cal{V}}_{k}}^{b} \,\Gamma_{IJ}^{ab} 
+ \frac1{54} \, \rho^{-2/3} \,  {\cal{V}}^{-1}{}^{gl} {{\cal{V}}_{k}}^{d}\,\varphi^{abc} \varphi^{efg} 
\,\Gamma_{IJ}^{abcdef}  \Big)  \nonumber\\
&&{} + \frac{1}{6}\, \rho^{-2/3} \frac{\partial {V}_{\mathrm{pot}}}{\partial \varphi^{abc}} \, \Gamma_{IJ}^{abc}
 \;,
\nonumber\\[1ex]
 0 &=&  -2 A_{IK} A_{KJ} + 2  \tilde{A}_{IK} \tilde{A}_{KJ} + 2 D_{IK} \tilde{B}_{KJ} + 2 \tilde{D}_{IK} B_{KJ} 
+ E_{KI}^{a} \tilde{C}_{KJ}^{a} - \tilde{E}_{KI}^{a} C_{KJ}^{a} \nonumber\\
&&{} - \delta_{IJ} \,   \frac{\partial {V}_{\mathrm{pot}}}{\partial \rho} 
- \frac{1}{3} \, \rho^{-1/3} \, {{\cal{V}}^{-1}}^{gl} {{\cal{V}}_{k}}^{a} \varphi^{bcg} \Gamma_{IJ}^{abc} \frac{\partial {V}_{\mathrm{pot}}}{\partial {Y_{k}}^{l}} \;,
\nonumber\\[1.5ex]
 0 &=& -2 C_{IK}^{a} A_{KJ} - 2 \tilde{C}_{IK}^{a} \tilde{A}_{KJ} + E_{IK}^{a} B_{KJ} + \tilde{E}_{IK}^{a} \tilde{B}_{KJ} 
+ 2 F_{IK}^{ab} C_{KJ}^{b} + 2 \tilde{F}_{IK}^{ab} \tilde{C}_{KJ}^{b}   \nonumber\\
&&{} - \rho^{-1}\, \frac{\partial {V}_{\mathrm{pot}} }{\partial {\Sigma^{ab}}} \Gamma_{IJ}^{b}
 - 3 \, \rho^{-1}\,\varphi^{bc(a}   \frac{\partial {V}_{\mathrm{pot}} }{\partial \varphi^{d)bc} } \Gamma_{IJ}^{d}
 - \frac{3}{2} \, \rho^{-1/3} \, {{\cal{V}}^{-1}}^{gl} {{\cal{V}}_{k}}^{[a} \varphi^{bc]g} \Gamma_{IJ}^{bc} \,
\frac{\partial {V}_{\mathrm{pot}} }{\partial {{Y_{k}}^{l}}} \;,
\nonumber\\[1ex]
 0 &=&  2 C_{IK}^{a} \tilde{A}_{KJ}  + 2 \tilde{C}_{IK}^{a} A_{KJ} + E_{IK}^{a} \tilde{B}_{KJ} + \tilde{E}_{IK}^{a} B_{KJ} 
+ 2 F_{IK}^{ab} \tilde{C}_{KJ}^{b} + 2 \tilde{F}_{IK}^{ab} C_{KJ}^{b}    
\nonumber\\[.5ex]
&&{} 
- \frac{\partial V_{\mathrm{pot}} }{\partial {{Y_{k}}^{l}}}   \,{{\cal{V}}_{k}}^{b}{\cal{V}}^{-1}{}^{c l} \,\Big( 
 \frac16 \, \rho^{-2/3} \,
 \Big(\varphi^{agh}\varphi^{efc} \delta^{db}-
 \delta^{b[a} \varphi^{gh]c}  \varphi^{def}\Big)  \Gamma_{IJ}^{defgh} 
 -\delta^{a(b} \, \Gamma_{IJ}^{c)}\Big) \nonumber\\
&&{} - \frac{3}{2} \, \rho^{-2/3} \, \frac{\partial V_{\mathrm{pot}} }{\partial \varphi^{abc} } \, \Gamma_{IJ}^{bc}  \;,
 \eea
where the last two equations should be understood as projected onto their gamma-traceless 
part in the indices $aI$. Remarkably, it turns out that all these equations are identically satisfied
for the solution (\ref{A-F}), (\ref{bc}) of the linear relations given in section~\ref{subsec:linear}.
This is a confirmation of the prediction of~\cite{Samtleben:2007an} discussed in section~\ref{subsec:embedding} above
that any embedding tensor of the type $\theta_{kl}$ automatically satisfies the relevant quadratic constraints
and thus defines a consistent gauged theory compatible with maximal supersymmetry.

\end{appendix}


\providecommand{\href}[2]{#2}\begingroup\raggedright\endgroup

\end{document}